\documentclass[article]{jss2}

\usepackage{amsmath,amssymb,latexsym}
\usepackage{subfig} % subfloating
\usepackage{enumerate}
\usepackage{color} 
\usepackage[dvipsnames]{xcolor}
\usepackage{booktabs} 

\usepackage{calc}  
\usepackage{enumitem}  
\usepackage{threeparttable}

\DeclareMathOperator{\Exp}{Exp}

\DeclareMathOperator{\PP}{\mathbf{P}}

\DeclareMathOperator{\uomega}{\underline{\omega}}

\DeclareMathOperator{\upi}{\underline\pi}

\graphicspath{{Figures/}} % subdirectory for figures

\author{Cristina Mollica\\MEMOTEF Department\\
 Sapienza University of Rome \And 
        Luca Tardella\\Department of Statistics\\
 Sapienza University of Rome}
\title{\pkg{PLMIX}: An \proglang{R} package for modeling and clustering partially ranked data}

\Plainauthor{Cristina Mollica, Luca Tardella} %% comma-separated
\Plaintitle{\pkg{PLMIX}: An R package for modeling and clustering partially ranked data} %% without formatting
\Shorttitle{\pkg{PLMIX}: An \proglang{R} package for modeling and clustering partially ranked data {\tiny ver. \today}} %% a short title (if necessary)

\Abstract{
Ranking data represent a peculiar form of multivariate ordinal data taking values in the set of permutations. 
Despite the numerous methodological contributions to increase the flexibility of ranked data modeling, the application of more sophisticated models is limited by the related computational issues.
The \pkg{PLMIX} package offers a comprehensive framework aimed at endowing the \proglang{R} statistical environment 
with some recent methodological advances
in modeling and clustering partially ranked data.
The usefulness of the novel \pkg{PLMIX} package can be motivated from several perspectives:
(i) it contributes to fill the gap concerning 
Bayesian estimation of ranking models in \proglang{R}, by focusing on the \textit{Plackett-Luce} model and its extension within the finite mixture approach as the generative sampling distribution;
(ii) it addresses computational complexity by combining the flexibility of \proglang{R} routines and the speed of compiled \proglang{C++} code, with possible parallel execution;
(iii) it covers the fundamental phases of ranking data analysis allowing for a more careful and critical application of ranking models in real experiments;
(iv) it provides effective tools for clustering heterogeneous partially ranked data. 
The functionality of the novel package is illustrated with several applications to simulated and real datasets.
}
\Plainkeywords{Ranking data, Plackett-Luce model, Mixture models, Bayesian inference, MAP estimation, MCMC methods, R} %% without formatting
\Address{
  Cristina Mollica\\
  MEMOTEF Department\\
  Sapienza University of Rome\\
  Piazzale A. Moro 5,\\
  00185 Rome, Italy\\
  E-mail: \email{cristina.mollica@uniroma1.it}\\
  
  Luca Tardella\\
  Department of Statistics\\
  Sapienza University of Rome\\
  Piazzale A. Moro 5,\\
  00185 Rome, Italy\\
  E-mail: \email{luca.tardella@uniroma1.it}
}
\begin{document}

%% include your article here, just as usual
%% Note that you should use the \pkg{}, \proglang{} and \code{} commands.

%\tableofcontents

\section{Introduction}
\label{s:intro}
%
%Let us consider an experiment in which a judge is ask to rank a set of $K$ labeled alternatives, namely \textit{items}, according to a certain criterion. The final outcome of the comparative evaluation is an ordered sequence of the given objects, called \textit{ranking}. 
A \textit{ranking} is an ordered sequence resulting from the comparative evaluation of a given set of \textit{items}.
This type of data 
%%% are 
is common in a large number of research fields, as testified by the 
% heterogenous 
widespread 
applications of ranking data analysis. 
%For example, in medicine physicians 
%%typically face with the situation of assessing the health status 
%often face with the problem to decide the most appropriate medical treatments or to plan clinical interventions from patients' qualitative perception on their health status
%% alternative causes of a specific symptom/manifestation or are asked to plan medical treatments and/or health interventions in order of priority
% ~\citep{Salomon,Krabbe2008,Ratcliffe2009}. 
% Moreover, 
 For example, various political election systems 
 %rely on voting systems such that 
 allow voters to express multiple preferences in the ballots rather than the only most-liked candidate
 %can 
%to be expressed 
~\citep{Stern1993,Gormley:Murphy-American},
%Gormley:Murphy-AnnalsApplied
implying that a method to aggregate the resulting rankings into a final \textit{consensus} is needed, as in~\cite{Davenport} and~\cite{Meila:Phadnis:al}.
%Bargagliotti. 
Furthermore, marketing and psychological studies are typically aimed at investigating individual preferences~\citep{Vigneau1999,Vitelli} and attitudes~\citep{Yu2005,Gormley:Murphy-Royal} towards multiple alternatives. 

In the biomedical context,~\cite{Mollica:Tardella} proposed the conversion of the quantitative
multivariate profiles resulting from a bioassay experiment into
ranking sequences. 
The ranking transformation was motivated as a
possible data normalization method 
when a well-established pre-processing technique is lacking.
%In their work, the ranking-based analysis was proved to provide a more robust evidence to successfully characterize the disease status.  

Moreover, sports and competitions 
%in general 
naturally
%contribute to 
motivate the development of methods for describing ranking outcomes, in order to quantify the ability of the competitors from the ordinal evidence, see for example~\cite{Henery-Royal},~\cite{Stern1990} and~\cite{Caron:Doucet}. Further references 
% pertaining each of the mentioned research fields 
can be found in the 
%introduction of 
the recent review
of the ranking literature supplied by~\cite{Alvo}. 
A public data repository devoted to 
preference data, although not necessarily in the form of rankings, is 
available at \url{http://www.preflib.org/} \citep{Mattei:Walsh:2013}.

From a mathematical perspective, ranking data represent a well-characterized form of multivariate ordinal data, with a direct correspondence with the set of permutations~\citep{Plackett1968,Stern1990}. 
By following the reviews in \cite{Critch:Flig:Verd},~\cite{Marden} and~\cite{Alvo}, 
%and the most recent monograph \cite{Alvo}, 
the main approaches 
%developed in the 
%ranking 
%literature 
to conceive 
%non-uniform 
parametric ranking models
% on $\mathcal{S}_K$ 
can be classified in four classes:
\begin{enumerate}
\item \textit{order statistics models} (OS), also known 
%in the ranking literature 
as random utility models,
% , where the 
whose cornerstone is 
% given by
  the \textit{Thurstone model} 
 %from the seminal work by 
%due to 
\citep{Thurstone};
% proposed by,
\item \textit{paired comparison models}, 
where the 
% whose 
%most general family is given by the Babington-Smith model described in~\cite{Smith};
most popular parametric family is the Bradley-Terry model (BT) described in~\cite{Bradley:Terry} and~\cite{Bradley84};
%, widely detailed in,
\item \textit{distance-based models} (DB), originally introduced by \cite{Mallows} and also referred to as Mallows' models;
%, introduced by,
\item \textit{stagewise models}.
%, described in.
\end{enumerate}
Each model class alludes to a specific generative process of the ordinal judgment 
%regarding 
on the given set of alternatives. 
%More specifically, 
This work concentrates on the last parametric family 
% moving from 
bearing on 
the decomposition of the ranking process into 
%$K-1$ 
%%% independent 
consecutive
stages, that is, the
%indicating
%corresponding to 
sequential selections 
%from the item set $I$.
of the 
%available 
items in order of preference.
In particular, our interest is in
%a
%very popular
%well-established
%parametric distribution,
%belonging to the stagewise family,
the \textit{Plackett-Luce model} (PL) and its finite mixture extension by assuming the Bayesian inferential perspective.
%, widely described in Section~\ref{ss:pl}.
% after a review of the 
%Bayesian methods for ranking models.

Despite the numerous methodological contributions of the last decades,
% aimed at increasing 
enhancing 
the flexibility of the aforementioned parametric classes, the application of more sophisticated ranking models is still limited in practice.
Likely, the main reason lies in the computational complexity emerging from the peculiar multivariate structure of ranking data, requiring the development of specialized software.
This might have
% has 
% often 
% traditionally 
% prevented from 
%precluded
slowed down a wider use of the most recent model proposals.
%Consequently, it turns out that inference of novel ranking models remains confined to methodological purposes rather than on the investigation of its usefulness in practice.
The \pkg{PLMIX} package version 1.0.1, released on CRAN (Comprehensive R Archive Network) and available at \url{ https://CRAN.R-project.org/package=PLMIX}, offers a comprehensive framework aimed at
%%%adjusting 
enriching
the \proglang{R} environment~\citep{Rsoft} 
%%%to 
with 
one of the recent methodological advances 
%%% achieved 
in modeling and clustering partially ranked data,
%ranked data analysis, 
by adequately accounting for the related computational issues.

The paper is organized as follows: Section~\ref{s:review} provides a detailed review of the existing packages
% is provided 
in \proglang{R} for ranking data and highlights the
main differences with the novel \pkg{PLMIX} package. The PL is briefly
reviewed in Section~\ref{s:pl} and its Bayesian extension to the finite mixture setting is detailed in Section~\ref{s:bayinf} with the related inferential procedures. Section~\ref{s:package} describes the application of the functions included in \pkg{PLMIX}
on simulated and real data examples.
%to perform a comprehensive Bayesian PL mixture analysis, %with the \pkg{PLMIX} package, 
%whereas illustrative applications follow in Section~\ref{s:examples}. 
The paper ends 
in Section~\ref{s:concl}
with some 
%%% final 
remarks and suggestions for future developments.

\section{Review of R packages for ranking data analysis}
\label{s:review}
\begin{table}[t]
\centering
\begin{threeparttable}
\caption{Characteristics of the existing \proglang{R} packages for ranking data compared with the novel \pkg{PLMIX} package.}
\renewcommand{\arraystretch}{1.2} % spreads   vertically
\begin{tabular}{lcccc}
%  \hline
Package & Ranking type & Model class & Mixture & Inference  \\ 
\hline
%\hline
%\pkg{ConsRank}         &   0 &   0 &   0 &   0  \\ 
\pkg{PerMallows}       &  Complete       &  DB, GMM  &  No   & MLE \\ 
\pkg{PlackettLuce}      &  Partial            &  PL  &  No   & MLE \\ 
\pkg{pmr}                   &  Complete       & DB,  WDB, PL    &   No   & MLE  \\ 
\pkg{prefmod}            &  Partial            &   BT                   &   Yes  & MLE  \\ 
\pkg{Rankcluster}      &  Partial            &  ISR                   &  Yes   & MLE  \\ 
\pkg{rankdist}            &  Partial            &   DB                   &  Yes   & MLE \\ 
\pkg{RMallow}           &  Partial            &  DB                   &  Yes   & MLE \\ 
\pkg{StatMethRank}  &  Complete       &  MNOS, WDB    & Yes   &  MLE$^*$  \\ 
\pkg{StatRank}          &  Partial            &  OS                    &  Yes   & MLE  \\ 
\hline
\pkg{PLMIX}            &  Partial  &   PL                    &  Yes   & Bayesian$^{**}$ \\ 
\hline
\end{tabular}
\label{t:pckchar}
\begin{tablenotes}
      \small
%      \item ISR $=$ Insertion Sort Rank model, GMM $=$ Generalized Mallows model, MNOS $=$ multivariate normal ordered statistic model, WDB $=$ weighted distance-based model  
      \item $^*$ a single function to fit the Bayesian MNOS is available
      \item $^{**}$ MLE can be recovered as special case under noninformative prior setting
\end{tablenotes}
\end{threeparttable}
\end{table}

Several \proglang{R} packages are currently available to conduct model-based
analysis of ranking data and their main features are summarized in
Table~\ref{t:pckchar}, in comparison with the novel \pkg{PLMIX}
package. Some 
% additional 
%%% detailed 
% details 
essential features 
% on 
of 
each package are provided in the following:
\begin{itemize}
%\item[-] \pkg{ConsRank} 
\item[-] \pkg{PerMallows}, described in \cite{Irurozki}, provides a suite of functions for the MLE of DBs and their multiparametric extensions, referred to as \textit{Generalized Mallows models} (GMM) in the seminal work by \cite{Fligner:Verducci-Royal}. Various metrics on the ranking space are considered, but partial rankings and finite mixtures are not contemplated;
\item[-] \pkg{PlackettLuce}, recently released on CRAN, performs ML inference of the PL from complete and partial rankings and includs methods to derive point estimates and standard errors even in critical situations when the likelihood function is not well-behaved. Additionally, the package can handle ties and admits the inclusion of covariates to accomplish a model-based partitioning of the sample units via PL trees. A full description of the package can be found in the vignette by \cite{PlackettLuce}; 
\item[-] \pkg{pmr}, presented in  \cite{Lee:Yu2013}, applies standard
MLE methods to infer several parametric ranking models, such as the
DB, 
%%% 
the \textit{weighted distance-based model} (WDB) proposed by \cite{Lee:Yu2010} and the PL. Ranking models are considered in their basic form (no mixture) and only complete rankings are allowed;
\item[-] \pkg{prefmod}, introduced in \cite{Hatz:Ditt:2012}, focuses on the analysis of preference data 
expressed in the form of 
%the equivalent representation of the ranking sequences into 
paired comparisons
% the form of 
(PC) and on the application of the BT and extensions thereof under the MLE approach.
%\textit{Bradley-Terry} model (BT). 
This package allows for the handling of partial observations, ties and the inclusion of individual and item-specific covariates. The generalization to
latent class settings 
%through a nonparametric MLE framework
%the finite mixture setting, 
is also possible via a nonparametric method, but it is limited to complete rankings;
\item[-] \pkg{Rankcluster}, 
%%% which 
widely described in \cite{Rankcluster}, implements the mixture of \textit{Insertion Sort Rank data models} (ISR), see \cite{Jacques2014}.
%, which extends the generative model for ranking data based on the Insertion Sort algorithm, originally defined in \cite{Biernacki:Jacques}, into the mixture approach. 
The ISR mixture is motivated as a model-based clustering tool of partial and potentially multivariate (hierarchical) ranking data;
\item[-] \pkg{rankdist}, based on the the methodological contribution by~\cite{Murphy:Martin}, fits mixtures of DBs with various metrics through the EM algorithm; it accepts both complete and partial rankings;
\item[-] \pkg{RMallow} 
%%% which 
implements the mixture of DBs with the Kendall distance as metric on the ranking space. Both complete and partial rankings are allowed;
\item[-] \pkg{StatMethRank} is in
%%% supporting 
support of
the
%
% recent 
monograph by \cite{Alvo}. 
%It mainly concerns exploratory techniques and (non mi viene). 
Regarding the parametric distributions,
% model-based analysis, 
it implements the mixture of WDBs from the MLE perspective and the Bayesian Multivariate Normal ordered statistics model (MNOS) described in \cite{Yu}, but exclusively on complete rankings;
% are not support for model-based inference. 
\item[-] \pkg{StatRank} 
%%% that 
covers the class of random utility models,
%better known in the ranking literature as ordered statistics models (OS), including the Plackett-Luce model 
involving the PL as special instance, and its generalization to the finite mixture context. 
% Frequentist e
Frequentist estimation is 
% achieved 
carried on by means of the Generalized Method-of-Moments~\citep{Soufiani2014} and can be performed also on partial observations.
%Likelihood of ? given ? under Gumbel is the same as the likelihood of ??, which is the inverse of ?, given ? under the exponential distribution. Therefore, P-L is equivalent to RUM with exponential distribution for the reverse profile, see also~\cite{Henery-Applied,Stern1990}.
\end{itemize}
%
%An
The 
%comparison 
outline
% of the
%with the 
%available \proglang{R} packages is illustrated 
in Table~\ref{t:pckchar}
%for ranking data 
points out that the existing
% contributed 
libraries cover a wide range
of the parametric options reviewed in Section~\ref{s:intro}. 
%Moreover, 
%%%  the most part of 
Most of 
them account also for the possible presence of incomplete observations and for the generalization of the ranking generative mechanism to the mixture framework. 

Nevertheless, with the only exception of the function
\code{mvnos.model} 
%included 
of the \pkg{StatMethRank} package implementing the Bayesian MNOS model on complete rankings via MCMC methods, all the available packages address inference from the frequentist point of view. Moreover, although \pkg{pmr} and \pkg{StatRank} encompass the PL distribution and its mixture extension, they either work only with complete observations or lack of computational efficiency, making sometimes prohibitive to perform a partial ranking analysis based on the PL mixture. The novelties introduced by the \pkg{PLMIX} package to overcome these limitations are widely described in Section~\ref{s:package}. An account of the methodological aspects implemented by \pkg{PLMIX} is provided in the next section. 

\section{The Plackett-Luce model for partial orderings}
\label{s:pl}

\subsection{Preliminaries and data format}
\label{ss:data}
%
%In this section we introduce notation and give basic definitions in
%the context of the ranked data theory.
Let us first clarify the basic terminology for the data input, in
particular the difference between 
ranking and ordering.
Formally, a \textit{full} (or \textit{complete}) \textit{ranking}
% rreevv
$\pi: I \to R$
is a bijective mapping of a finite set
$I=\{1,\dots,K\}$ of labeled \textit{items} (or alternatives) into a set of \textit{ranks} $R=\{1,\dots,K\}$,
resulting from the attribution of a position to each item according to a determined criterion.
%Hence, $K$ is the total number of items to be ranked and 
The result of 
the mapping
can be represented in terms of 
the 
%an ordered 
$K$-tuple $\pi=(\pi(1),\dots,\pi(K))$, where the generic entry $\pi(i)$ indicates the rank assigned to the $i$-th item.
%According to the underlying convention in the ranking theory, 
If $\pi(i)<\pi(i')$, then item $i$ is said to be ranked higher than/preferred to item $i'$. 
%In the psychological literature ranked sequences are also referred to as \textit{ordinal ipsative data} because of the constant sum of their components \citep{Chan:Bent}.

%Formally, a \textit{full} (or \textit{complete}) \textit{ranking} is a bijective mapping of a set $I$ of labeled \textit{items} into a set of \textit{ranks} $R$, that is
%$$\pi: I \to R,$$
%where for simplicity we identify both finite sets with the first $K$ positive integers $\{1,\dotsc,K\}$.

Ranking data admit an alternative format in terms of orderings. Specifically,
the  \textit{full} (or \textit{complete}) \textit{ordering} $\pi^{-1}: R \to I$ is simply the inverse function of the ranking $\pi$, yielding the ordered vector $\pi^{-1}=(\pi^{-1}(1),\dotsc,\pi^{-1}(K))$ whose generic component $\pi^{-1}(j)$ denotes the item ranked in the $j$-th position.

In many real applications, for example when $K$ is large, 
%it often happens
% it is possible
%that 
the ranking elicitation could be not completely carried out. A typical
situation is when the ranker specifies only her most-liked $t<K$ items
and leaves the remaining $K-t$ positions undefined. In this case, the
generic observation consists in the so-called \textit{top-$t$ partial
  ordering} of the form
$\pi^{-1}=(\pi^{-1}(1),\dots,\pi^{-1}(t))$. With a slight abuse of
notation, the remaining $K-t$ alternatives are tacitly assumed to be
ranked lower, formally $\pi(i)>t$ for all
$i\notin\{\pi^{-1}(1),\dots,\pi^{-1}(t)\}$. Notice that a complete
ordering is a special instance of top-$t$ partial ordering with
$t=K-1$,
%In fact, a top-$(K-1)$ sequence
%%In this regard, that the case $t=K-1$ 
%corresponds to a complete ordering, 
since the single missing 
$K$-th entry 
can be unambiguously determined.  
Finally, we remark that in the present context ties, i.e., the case when multiple items occupy the same position,
% in the ranking, 
are not contemplated.

\subsection{The Plackett-Luce model}
\label{ss:pl}
The PL is 
%one of the most popular and frequently applied 
one of the most successfully applied
\textit{stagewise models}
% parametric distribution 
to describe partially ranked data, whose paternity is jointly attributed to~\cite{Luce} and~\cite{Plackett}.
%in the context of model-based ranking data analysis.
%of partial rankings 
%It belongs to the class of \textit{stagewise ranking models}, where
%% of a finite item set.
%the elicitation of the comparative judgment on the $K$ alternatives
%% expressed in ordinal form, in fact, 
%is 
%%conceived as a sequential process 
%divided into $K-1$ stages:
%%% 
The ranking elicitation is conceived as a random sampling without replacement from an urn:
at each stage the most-liked item is specified among 
%%% those 
the 
alternatives 
not selected at the previous stages. The sequential draws of the items are governed by the \textit{support parameters} $\underline{p}=(p_1,\dots,p_K)$, that is, positive constants representing a measure of liking toward each item.
%More specifically, 
Let $\pi^{-1}_s=(\pi^{-1}_s(1),\dots,\pi^{-1}_s(n_s))$ be a generic top partial ordering, where $n_s$ is the number of items ranked by unit $s$ in the first $n_s$ positions. The PL postulates 
%the following probability for the generic partial observation $\pi^{-1}_s$ 
%
\begin{equation}
\label{e:pl}
\PP_{\text{PL}}(\pi^{-1}_s|\underline{p})=\prod_{t=1}^{n_s}\dfrac{p_{\pi_s^{-1}(t)}}{\sum_{i=1}^{K}p_i-\sum_{\nu=1}^{t-1}p_{\pi_s^{-1}(\nu)}}.
\end{equation}
%
%As apparent in~\eqref{e:pl}, the underlying assumption is that the relative preference of each item over the other ones is constant across the ranking stages.
%Since the present MLE procedure described in~\cite{Caron:Doucet} is based on the direct extension of the BT, only for this specific inferential framework item parameters $\underline{p}$ are intended to be unknown positive quantities, not formally subject to the constraint $\sum_{i=1}^Kp_i=1$; as these are identified up to the multiplication by a positive constant, we continue to count $K-1$ free parameters.
%Moreover, for completeness we stress that model for multiple comparisons presented in~\cite{Caron:Doucet} aims at accounting also for the possible presence of top-$t$ partial rankings but the normalization of the $p$'s at each stage is $s$-dependent, in the sense that it is taken w.r.t.\ the items actually ranked in every specific ordinal sequence. This is different, for example, from the PL for the same type of incomplete rankings detailed in the series of works by I. C. Gormley and T. B. Murphy, where the normalization is constant over observed rankings and contemplates the whole item set $I$. Although it is useful to known and  keep this difference in mind, we do not have to worry about it in
%% the present
%applications
%% because we observed only
%limited to full rankings.
%, therefore the PL formulation found in~\cite{Caron:Doucet} reduces to~\eqref{e:PL} previously introduced.
%It follows that, 
For a given a random sample $\underline{\pi}^{-1}=\{\pi_s^{-1}\}_{s=1}^N$ of $N$ partial top orderings with varying lenghts, the observed-data log-likelihood turns out to be
\begin{equation}
\label{e:plobsloglik}
l(\underline{p})=\sum_{i=1}^K\gamma_i\log{p_i}-\sum_{s=1}^N\sum_{t=1}^{n_s}\log\sum_{i=1}^K\delta_{sti}p_i,
\end{equation}
where $\gamma_i=\sum_{s=1}^Nu_{si}$ with $u_{si}=\mathbb{I}_{[i\in\{\pi_s^{-1}(1),\dots,\pi_s^{-1}(n_s)\}]}$ and $\delta_{sti}=\mathbb{I}_{[i\notin\{\pi_s^{-1}(1),\dots,\pi_s^{-1}(t-1)\}]}$ with $\delta_{s1i}=1$ for all $s=1,\dots,N$ and $i=1,\dots,K$.

\section{The Bayesian Plackett-Luce mixture model}
\label{s:bayinf}

In this section we give a brief outline of the Bayesian approach based on the data augmentation strategy to make inference on the PL parameters, both in the case of homogeneous population without an underlying group structure and in the more general finite mixture framework. It represents the methodological background implemented in the \pkg{PLMIX} package.

\subsection{The homogeneous case}
\label{ss:homo}

Because of the normalization term $\sum_{i=1}^K\delta_{sti}p_i$, the direct maximization of the log-likelihood~\eqref{e:plobsloglik} is not straightforward. In the Bayesian setting,
simple and effective estimation 
procedures
%for the PL 
%described in Section~\ref{ss:pl} 
were introduced by \cite{Caron:Doucet} to overcome this inconvenience.
% form of the PL likelihood.
%In their work the PL parametric approach is referred to as a model for multiple comparisons, to stress the contrast with the pairwise comparisons where $K=2$. 
Their crucial idea 
%due to the normalization term $\sum_{i=1}^K\delta_{sti}p_i$ 
relies on a data augmentation step with continuous latent variables 
%in the model specification.
associated to each entry of the observed matrix.
More specifically,
%Their data augmentation method consists in the introduction of a suitable set of continuous latent variables $\underline{Y}=(Y_{st})$ for $s=1,\dots,N$ and $t=1,\dots,n_s$, associated to each entry of the observed matrix. leading to an EM algorithm equivalent to the Minimization/Maximization (MM)
%%algorithm
%technique as
%proposed by~\cite{Hunter}.
%Thus,~\cite{Caron:Doucet} contributed to make more explicit why the EM algorithm
%is a special case of the wider class of MM algorithms,
%as demonstrated by~\cite{Heiser}, allowing to possibly exploit methods
%to accelerate the former procedure and make it more efficient.
\cite{Caron:Doucet} suggest to employ
%associate to each entry of the observed data matrix
% the
%complete the sampling space with 
auxiliary variables $\underline{y}=(y_{st})$ for $s=1,\dots,N$ and $t=1,\dots,n_s$
with a suitable parametric assumption for their joint conditional distribution,
%,that is,
given by
\begin{equation}
\label{e:fullcond}
f(\underline{y}|\upi^{-1},\underline{p})=\prod_{s=1}^N\prod_{t=1}^{n_s}f_{\Exp}\biggl(y_{st}\biggr\vert\sum_{i=1}^K\delta_{sti}p_i\biggr),
\end{equation}
where $f_{\Exp}(\cdot|\lambda)$ is the Negative Exponential density
function
%expressed in terms of
indexed by the rate parameter
$\lambda$.
%The parametric assumption \eqref{e:fullcond} does not alter the marginal structure of the model but, as detailed shortly, entails decisive facilitations
%% from an analytic point of view.
%in deriving closed-forms for both the optimization and the GS algorithm.
%%introduce
%%latent variables
%%are assumed
%In this regard, notice that the rate parameters $\lambda_{st}=\sum_{\nu=t}^{K}p_{\pi_s^{-1}(\nu)}$ of the $Y$'s correspond to the sequential normalizations of the $p$'s, that is the annoying terms of the PL likelihood.
Additionally, assumption~\eqref{e:fullcond}
is conveniently combined with a conjugate prior distribution $f_0(\underline{p})=\prod_{i=1}^Kf_{\text{Ga}}(p_{i}|c,d)$ for the support parameters, where $c$ and $d$ denote the shape and rate parameters of the Gamma densities, leading to a straightforward Bayesian inference.
%which clearly suggests the adoption of 
%independent Gamma priors for the support parameters %$p_i\sim\text{Ga}(c,d)$, where $c$ and $d$ are respectively the shape and rate parameter, 
%$\underline{p}$ to obtain a conjugate model specification.

\subsubsection{MAP estimation via EM algorithm}
\label{s:MAPhomo}
%
%As first type of inference, \cite{Caron:Doucet} describe how to achieve the Maximum A Posteriori (MAP) estimate of the vector $\underline{p}$, i.e., the posterior mode.
In the presence of latent variables, the popular EM algorithm introduced by~\cite{Demp:Lai:Rub} can be applied to optimize the posterior distribution and achieve the Maximum A Posteriori (MAP) estimate of
% the vector $\underline{p}$, 
the PL parameters, i.e., the posterior mode.
At the generic iteration $l+1$, the EM algorithm described by~\cite{Caron:Doucet} updates the support parameters as follows
\begin{equation*}
\label{}
p_i^{(l+1)}=\dfrac{c-1+\gamma_i}{d+\sum_{s=1}^N\sum_{t=1}^{n_s}\dfrac{\delta_{sti}}{\sum_{i=1}^K\delta_{sti}p_i^{(l)}}}\qquad i=1,\dots, K.
\end{equation*}
%where
%%
%\begin{equation*}
%\delta_{sti}=\begin{cases}
%      1\qquad\text{ if }i\in\{\pi_s^{-1}(t),\dots,\pi_s^{-1}(n_s)\},\\
%      0\qquad\text{ otherwise}
%\end{cases}
%\end{equation*}
%%
%and $\gamma_i=\sum_{s=1}^Nu_{si}$ with
%%
%\begin{equation*}
%u_{si}=\begin{cases}
%      1\qquad\text{ if }i\in\{\pi_s^{-1}(1),\dots,\pi_s^{-1}(n_s)\},\\
%      0\qquad\text{ otherwise}.
%\end{cases}
%\end{equation*}
%%
%The quantity $\gamma_i$ denotes the number of sample units who assigned a position to item $i$, whereas $\delta_{sti}$ is the indicator that item $i$ appears in a position not better than $t$ in the $s$-th partial ranking.
By setting noninformative hyperparameters $c=1$ and $d=0$, the EM procedure reduces to the 
%minorizing auxiliary objective function of the 
%iterative 
Minorization-Maximization algorithm 
%(MM) 
described by~\cite{Hunter}
% to 
%%%% perform 
%compute
for the MLE of the PL.
%It is possible to update current parameter values in a cyclic way, i.e., to compute $p_i^{(l+1)}$ employing the most recent available values for the $p$'s. Conditions under which the sequence of updates $\underline{p}^{(1)},\underline{p}^{(2)},\dots$ maximizing~\eqref{e:Q} is guaranteed to converge to the ML estimates are given in~\cite{Hunter}, together with an argument
%%of
%on
%their validity in this case of multiple comparisons.
%We
%%have to notify that
%briefly mention that \cite{Caron:Doucet}
%developed a more general inferential method employing
%independent Gamma priors for the components of $\underline{p}$
%and deriving the MM algorithm for MLE estimation
%as a
%%special case
%by-product
%of the Bayesian approach.
%Hence,
%our notation for the M-step optimum can be easily obtained from their work setting noninformative hyperparameters values, i.e., $a=1$ and $b=0$
%respectively for the shape and the rate parameter of the prior distribution.
%To implement a fully Bayesian inferential method,

\subsubsection{Gibbs sampling}
\label{s:GShomo}

\cite{Caron:Doucet} describe also the Gibbs sampling (GS) procedure, that is, a 
%%% sampling-based 
simulation-based 
method to approximate the joint posterior distribution 
%from the 
%joint 
%posterior distribution 
and to
%learn about the uncertainty associated to the final estimates
assess the uncertainty of the parameter estimates with empirical summaries of posterior variability.
%The algorithm requires the identification of the \textit{full-conditional distributions}, which are the distributions of a subset of the unobserved variables given all the remaining unknown quantities and the data.
%In this regard, first notice that the full-conditional of $\underline{y}$
%%determined by construction 
%is imposed by the data augmentation assumption~\eqref{e:fullcond}.
%Finally, thanks to the conjugate 
%%structure
%prior specification, the full-conditionals of the support parameters
%still belong to the Gamma 
%%%% parametric 
%family. 

%Specifically,
%%
%\begin{equation*}
%%\begin{split}
%\PP(p_i|\upi^{-1},\underline{y},p_{[-i]})\propto
%f_0(p_i)L_c(\underline{p},\underline{y})
%%&\propto p_i^{c-1}e^{-dp_i}\prod_{s=1}^Np_i^{u_{si}}e^{-p_i\sum_{t=1}^{n_s}\delta_{sti}y_{st}}\\
%=p_i^{c+\gamma_i-1}e^{-p_i(d+\sum_{s=1}^N\sum_{t=1}^{n_s}\delta_{sti}y_{st})},
%%\end{split}
%\end{equation*}
%%
%where $p_{[-i]}$ denotes the vector $\underline{p}$ of the support parameters without the $i$-th component.
%In conclusion, 
At the generic iteration $l+1$, the GS alternates 
%iteratively 
%by alternating 
the following two sampling steps
\begin{eqnarray*}
%\item[-] for $s=1,\dots,N$ and $t=1,\dots,n_s$ sample
y_{st}^{(l+1)}|\pi_s^{-1},\underline{p}^{(l)} & \sim & \Exp\left(\sum_{i=1}^K\delta_{sti}p_i^{(l)}\right),\\
%\item[-] for $i=1,\dots,K$ sample
p_{i}^{(l+1)}|\underline\pi^{-1},\underline{y}^{(l+1)} & \sim & \text{Ga}\left(c+\gamma_{i},d+\sum_{s=1}^{N}\sum_{t=1}^{n_s}\delta_{sti}y_{st}^{(l+1)}\right),
\end{eqnarray*}
where
the full-conditional of $\underline{y}$
%determined by construction 
is imposed by the data augmentation assumption~\eqref{e:fullcond} and
the full-conditionals of the $p$'s
%support parameters
belong to the Gamma 
%%% parametric 
family, thanks to the conjugate 
%structure
prior specification.

\subsection{The finite PL mixture}
\label{ss:hetero}

We now review the proposal recently developed by \cite{Mollica:Tardella2017} to extend the data augmentation approach~\eqref{e:fullcond}
% for the homogeneous PL 
%detailed in the subsection~\ref{s:homo} 
to the finite mixture context.

Formally, the $G$-component PL mixture model assumes that observations are sampled from a heterogeneous population composed of $G$ subpopulations called \textit{mixture components}
%for $g=1,\dots,G$, 
%such that
%that is,
%
\begin{equation}
\label{e:mpl}
\pi_s^{-1}|\underline{p},\uomega\,\overset{\text{iid}}{\sim}\,\sum_{g=1}^G\omega_g\PP_{\text{PL}}(\pi^{-1}_s|\underline{p}_g),%\qquad s=1,\dots,N,
\end{equation}
where  each component $g$ follows a basic PL distribution with a specific support parameter vector $\underline{p}_g$ and $\uomega=(\omega_1,\dots,\omega_G)$ are
% referred to as 
the \textit{mixture weights}.
% associated to the PL components.
Let $\underline{z}_s=(z_{s1},\dots,z_{sG})|\uomega\sim\text{Multinom}(1,\uomega=(\omega_1,\dots,\omega_G))$ be the vector describing the latent group membership of unit $s$,
%whose generic entry is
such that
\begin{equation*}
z_{sg}=\begin{cases}
      1\qquad \text{if unit $s$ belongs to the $g$-th mixture component}, \\
      0\qquad \text{otherwise}.
\end{cases}
\end{equation*}
%
%As apparent, the parametric representation~\eqref{e:mpl} defines a latent variable model where the unobserved group configuration is marginalized out. 
%suggest to account for the latent group structure in the data augmentation step as follows
To account for the latent group structure $\underline{z}$, \cite{Mollica:Tardella2017} generalize~\cite{Caron:Doucet}'s approach 
with the following conjugate Bayesian model setup
%%
%\begin{align*}
%\uomega & \sim  \text{Dir}(\alpha_1,\dots,\alpha_G)\\
%p_{gi} & \overset{i}{\sim}  \text{Ga}(c_{gi},d_g) \\ %\qquad g=1,\dots,G \text{and} i=1,\dots,K
%\underline{z}_s|\uomega & \overset{iid}{\sim} \text{Multinom}(1,\uomega)\\
%\pi_s^{-1}|\underline{z}_s,\underline{p} & \overset{i}{\sim}  \prod_{g=1}^G\PP_{\text{PL}}(\pi^{-1}_s|\underline{p}_g)^{z_{sg}} \\ %\qquad s=1,\dots,N
%y_{st}|\pi_s^{-1},\underline{z}_s,\underline{p}  & \overset{i}{\sim}  \text{Exp}\left(\prod_{g=1}^G\left(\sum_{i=1}^K\delta_{sti}p_{gi}\right)^{z_{sg}}\right).  %\qquad s=1,\dots,N \text{and} t=1,\dots,n_s
%\end{align*}
%%
%
\begin{eqnarray*}
\uomega & \sim & \text{Dir}(\alpha_1,\dots,\alpha_G)\\
p_{gi} & \overset{\text{i}}{\sim} & \text{Ga}(c_{gi},d_g) \\ %\qquad g=1,\dots,G \text{and} i=1,\dots,K
\underline{z}_s|\uomega & \overset{\text{iid}}{\sim} & \text{Multinom}(1,\uomega)\\
\pi_s^{-1}|\underline{z}_s,\underline{p} & \overset{\text{i}}{\sim} & \prod_{g=1}^G\PP_{\text{PL}}(\pi^{-1}_s|\underline{p}_g)^{z_{sg}} \\ %\qquad s=1,\dots,N
y_{st}|\pi_s^{-1},\underline{z}_s,\underline{p}  & \overset{\text{i}}{\sim} & \text{Exp}\left(\prod_{g=1}^G\left(\sum_{i=1}^K\delta_{sti}p_{gi}\right)^{z_{sg}}\right).  %\qquad s=1,\dots,N \text{and} t=1,\dots,n_s
\end{eqnarray*}

\subsubsection{MAP estimation via EM algorithm}
\label{sss:MAPhete}

In the mixture setting, the $(l+1)$-th 
%generic 
iteration of the EM algorithm
consists in updating
%alternates the E-step, which requires the computation of 
%%the expected log-posterior distribution 
%$Q((\underline{p},\uomega),(\underline{p}^*,\uomega^*))
%=\mathbb{E}_{\underline{y},\underline{z}|\upi^{-1},\underline{p}^*,\uomega^*}[l_c(\underline{p},\uomega,\underline{y},\underline{z})]+\log f_0(\underline{p},\uomega)$ w.r.t. the conditional distribution of all the latent variables, 
%%$\PP(\underline{y},\underline{z}|\upi^{-1},\underline{p},\uomega)=f(\underline{y}|\upi^{-1},\underline{z},\underline{p},\uomega)
%%\PP(\underline{z}|\upi^{-1},\underline{p},\uomega)$, whose maximization 
%and the subsequent 
%%maximization
%M-step, yielding 
%the following update rules of 
the unknown quantities until convergence according to the following formulas
%for the M-step 

%%The M-step of the EM reduces to iterate until convergence :\\
%%%
%%\begin{itemize}
%%\item[] \textbf{Initialization:} %\hspace{5 mm}]
%%set starting values $\underline{p}^{(0)}, \underline\omega^{(0)}$ for the parameters to be estimated;\\
%%%\vspace{4 mm}
%%\item[] \textbf{Computation:} %\hspace{5 mm}]
%%at iteration $l+1$, compute until convergence
%%\vspace{2 mm}
%\begin{itemize}
%\item[-] for $s=1,\dots,N$ and $g=1,\dots,G$ compute
%$$ \hat z_{sg}^{(l+1)}=\dfrac{\omega_g^{(l)}\PP_{\text{PL}}(\pi^{-1}_s|\underline{p}_g^{(l)})}{\sum_{g'=1}^G\omega_{g'}^{(l)}\PP_{\text{PL}}(\pi^{-1}_s|\underline{p}_{g'}^{(l)})},$$
%\item[-] for $g=1,\dots,G$ compute
%$$\omega_g^{(l+1)}=\dfrac{\alpha_g-1+\sum_{s=1}^N \hat z_{sg}^{(l+1)}}{\sum_{g'=1}^G\alpha_{g'}-G+N},$$
%\item[-] for $g=1,\dots,G$ and $i=1,\dots,K$ compute
%$$p_{gi}^{(l+1)}=\dfrac{c_{gi}-1+\hat\gamma_{gi}^{(l+1)}}{d_g+\sum_{s=1}^N \hat z_{sg}^{(l+1)}\sum_{t=1}^{n_s}\dfrac{\delta_{sti}}{\sum_{\nu=t}^{K}p_{g\pi_s^{-1}(\nu)}^{(l)}}},$$
%where $\hat\gamma_{gi}^{(l+1)}=\sum_{s=1}^N \hat z_{sg}^{(l+1)}u_{si}$.
%\end{itemize}
%%\end{itemize}
%%
%The M-step of the EM reduces to iterate until convergence :\\
%%
%\begin{itemize}
%\item[] \textbf{Initialization:} %\hspace{5 mm}]
%set starting values $\underline{p}^{(0)}, \underline\omega^{(0)}$ for the parameters to be estimated;\\
%%\vspace{4 mm}
%\item[] \textbf{Computation:} %\hspace{5 mm}]
%at iteration $l+1$, compute until convergence
%\vspace{2 mm}
\begin{align*}
%\item[-] for $s=1,\dots,N$ and $g=1,\dots,G$ compute
\hat z_{sg}^{(l+1)}&=\dfrac{\omega_g^{(l)}\PP_{\text{PL}}(\pi^{-1}_s|\underline{p}_g^{(l)})}{\sum_{g'=1}^G\omega_{g'}^{(l)}\PP_{\text{PL}}(\pi^{-1}_s|\underline{p}_{g'}^{(l)})},\\[10pt]
%\item[-] for $g=1,\dots,G$ compute
\omega_g^{(l+1)}&=\dfrac{\alpha_g-1+\sum_{s=1}^N \hat z_{sg}^{(l+1)}}{\sum_{g'=1}^G\alpha_{g'}-G+N},\\[10pt]
%\item[-] for $g=1,\dots,G$ and $i=1,\dots,K$ compute
p_{gi}^{(l+1)}&=\dfrac{c_{gi}-1+\hat\gamma_{gi}^{(l+1)}}{d_g+\sum_{s=1}^N \hat z_{sg}^{(l+1)}\sum_{t=1}^{n_s}\dfrac{\delta_{sti}}{\sum_{i=1}^{K}\delta_{sti}p_{gi}^{(l)}}},
\end{align*}
%\end{itemize}
%
where $\hat\gamma_{gi}^{(l+1)}=\sum_{s=1}^N \hat z_{sg}^{(l+1)}u_{si}$.
Interestingly, under the 
%%% uninformative
noninformative prior setting 
%corresponding to 
($c_{gi}=1$, $d_g=0$ and $\alpha_g=1$), the above MAP procedure recovers the MLE method to infer the PL mixture described by \cite{Gormley:Murphy-Royal}. 

\subsubsection{Gibbs sampling}
\label{sss:GShete}

Thanks to the conjugate prior specification, all the full-conditional distributions have known form and are easy to be sampled. 
%With very simple algebra, see for more details \cite{Mollica:Tardella2017}, one can easily verify that, 
At the generic iteration $l+1$,
the GS algorithm 
%to approximate the joint posterior distribution $\PP(\underline{z},\underline{y},\underline{p},\underline\omega|\underline\pi^{-1})$ 
consists in iteratively generating random values
%for all the unobserved quantities 
from the following full-conditionals
\begin{eqnarray*}
\underline\omega^{(l+1)}|\underline{z}^{(l)} & \sim & \text{Dir}\left(\alpha_1+\sum_{s=1}^Nz_{s1}^{(l)},\dots,\alpha_G+\sum_{s=1}^Nz_{sG}^{(l)}\right)\\[10pt]
y_{st}^{(l+1)}|\pi_s^{-1},\underline{z}_s^{(l)},\underline{p}^{(l)} & \sim & \text{Exp}\left(\prod_{g=1}^G\left(\sum_{i=1}^{K}\delta_{sti}p_{gi}^{(l)}\right)^{z_{sg}^{(l)}}\right)\\[10pt]
p_{gi}^{(l+1)}|\underline\pi^{-1},\underline{y}^{(l+1)},\underline{z}^{(l)} & \sim & \text{Gam}\left(c_{gi}+\gamma_{gi}^{(l)},d_g+\sum_{s=1}^Nz_{sg}^{(l)}\sum_{t=1}^{n_s}\delta_{sti}y_{st}^{(l+1)}\right)\\[10pt] %\qquad g=1,\dots,G \text{and} i=1,\dots,K
\underline{z}_s^{(l+1)}|\pi_s^{-1},\underline{y}_s^{(l+1)},\underline{p}^{(l+1)},\underline\omega^{(l+1)} & 
\sim & \text{Multinom}\left(1,\left(m_{s1}^{(l+1)},\dots,m_{sG}^{(l+1)}\right)\right),%\qquad s=1,\dots,N and g=1,\dots,G
\end{eqnarray*}
where $\gamma_{gi}^{(l)}=\sum_{s=1}^N z_{sg}^{(l)}u_{si}$ and
%$$m_{sg}^{(l+1)}=\dfrac{\omega_g^{(l+1)}\prod_{i=1}^K(p_{gi}^{(l+1)})^{u_{si}}
%e^{-p_{gi}^{(l+1)}\sum_{t=1}^{n_s}\delta_{sti}y_{st}^{(l+1)}}}
%{\sum_{g'=1}^G\omega_{g'}^{(l+1)}\prod_{i=1}^K(p_{g'i}^{(l+1)})^{u_{si}}e^{-p_{g'i}^{(l+1)}\sum_{t=1}^{n_s}\delta_{sti}y_{st}^{(l+1)}}},$$
$m_{sg}^{(l+1)}\propto\omega_g^{(l+1)}\prod_{i=1}^K(p_{gi}^{(l+1)})^{u_{si}}
e^{-p_{gi}^{(l+1)}\sum_{t=1}^{n_s}\delta_{sti}y_{st}^{(l+1)}},$
see \cite{Mollica:Tardella2017} for more analytical details.
The MAP solution represents a suitable starting point to
initialize the GS algorithm. 
%% recalled 
%as better discussed 
%%illustrated 
%in the next section.
%% can be recovered.

\subsubsection{Label-switching issue}
\label{sss:LS}
%

%When one adopts an MCMC simulation to derive approximate Bayesian inference of a mixture model,
%%from the posterior distribution of the complete model representation,
%an annoying identifiability issue can affect the posterior sample: the \textit{label switching} (LS).
The \textit{label switching} (LS) is an identifiability issue that can hamper the straightforward use of the MCMC simulations for the Bayesian estimation of mixture models \citep{Marin:Mengersen:Robert}.
%The LS  
It
%represents a specific form of unidentifiability concerning mixture models, that
% is
reflects the arbitrary attribution of the indices $\{1,\dots,G\}$ to denote the mixture components, 
%This means 
such that the
%The application of a permutation $\xi\in\mathcal{S}_G$ of the $G$ indices to a given parameter point, which corresponds to a 
relabeling of the latent classes does not modify the resulting sampling distribution.
To solve the LS problem
%MCMC analysis of mixture models,
in the GS output,
%, see 
%\citep{Jasra},
%for a review,
%Following the review by \cite{Jasra}, they can be summarized in three classes: (i) introduction of artificial identifiability constraints, (ii) relabeling algorithms (RA) and (iii) employment of label-invariant loss functions.
%The first approach consists in the elicitation of restrictions over the parameter space, typically order relations, which are satisfied by only one labeling $\tau$ of the mixture components.
%% and
%This action forces the equivalence classes to be singular so that LS ambiguity no longer persists. The pioneering work in this direction was \cite{Diebolt:Robert}, followed by the application of identifiability constraints in \cite{Richardson:Green}. Practical implications and criticisms related to this method are reviewed and discussed in \cite{Marin:Robert} and \cite{Jasra}. We simply note that in our multivariate setting the specification of artificial constraints can be very arduous.
we focus
%%% focused
%concentrate our attention
on the relabeling algorithms (RA), where the basic idea
% of the RA
%is the post-processing, that is
is the \textit{ex-post} relabeling of the raw MCMC samples in order to derive
%make them lie in a unique posterior mode among the $G!$ possible modal regions. 
meaningful posterior estimates.
A comprehensive review
% of the RA 
can be found in~\cite{Papastamoulis},
describing their implementation in the
% recently released 
\proglang{R}
package \pkg{label.switching}, that we exploited
% for the treatment of 
to handle the LS in our Bayesian PL mixture applications.
% provided in Section~\ref{s:examples}. 

%We conclude this brief review of the solutions for the LS mentioning the strategy based on the decision theoretic approach, where meaningful Bayesian estimates in presence of LS are obtained with the minimization of the posterior expectation of label-invariant loss functions. We do not go into further details of such an approach but the interested reader can refer to the fundamental works by \cite{Celeux:Hurn:Robert} and \cite{Hurn:Justel:Robert} for both the theoretical and practical aspects related to the application of the label-invariant loss functions.

\subsection{Bayesian model comparison criteria}
\label{ss:mc}

A crucial step in the finite mixture analysis 
%described
% in Section~\ref{ss:hetero} 
is the determination of the optimal number $\hat G$ of components
that, in general, is not known \textit{a priori}. 

%
%%%
%\begin{table}[t]
%\caption{Model selection criteria implemented in the \pkg{PLMIX} package.}
%\small% smaller size
%%\addtolength{\tabcolsep}{-3pt}    % shrinks horizontally
%\renewcommand{\arraystretch}{1.8} % spreads   vertically
%%\sideways
%\label{t:Bayselection}
%%\begin{center}
%\centering
%\begin{tabular}
% {llllll} 
%%\multicolumn{2}{c}{Model selection criteria} \\
%Criterion & Formula & & & Criterion & Formula \\
%\hline
%DIC$_1$ & $\bar D + (\bar D-D(\hat\theta_{\text{MAP}}))$ & & & BICM$_1$ & $\bar D + \frac{\mathbb{VAR}[D(\theta)|\underline\pi^{-1}]}{2}(\log N -1)$ \\
%DIC$_2$ & $\bar D + \dfrac{\mathbb{VAR}[D(\theta)|\underline\pi^{-1}]}{2}$ & & & BICM$_2$ & $D(\hat\theta_{\text{MAP}}) + \frac{\mathbb{VAR}[D(\theta)|\underline\pi^{-1}]}{2}\log N$ \\
%DIC$_3$ & $D(\hat\theta_{\text{MAP}}) + (\bar D-D(\hat\theta_{\text{MAP}}))\log N$ & & & WAIC$_1$ & $-2(lppd-p_{\text{WAIC$_1$}})$   \\
%BPIC$_1$ & $\bar D + 2(\bar D-D(\hat\theta_{\text{MAP}}))$ & & & WAIC$_2$ & $-2(lppd-p_{\text{WAIC$_2$}})$ \\
%BPIC$_2$ & $\bar D + 2\mathbb{VAR}[D(\theta)|\underline\pi^{-1}]$ & & & MOB & $\underset{G}{\text{argmax}\;} W^+_G$\\
%BIC & $D(\hat\theta_{\text{MAP}}) + k\log N$ & & & SMD & $\underset{G}{\text{argmin}\;} \mathbb{MED}[D_G(\theta)|\underline\pi^{-1}]$ \\
%\hline
%\end{tabular}
%%\end{center}
%%\endsideways
%\end{table}
%%

%
\begin{table}[t]
\caption{Model selection criteria implemented in the \pkg{PLMIX} package.}
%\small% smaller size
%\addtolength{\tabcolsep}{-3pt}    % shrinks horizontally
\renewcommand{\arraystretch}{1.8} % spreads   vertically
%\sideways
\label{t:Bayselection}
%\begin{center}
\centering
\begin{tabular}
 {ccccc} 
%\multicolumn{2}{c}{Model selection criteria} \\
\textbf{DIC}$_\textbf{1}$ & \quad & \textbf{BPIC}$_\textbf{1}$ & \quad & \textbf{BICM}$_\textbf{1}$ \\
\hline
$\bar D + (\bar D-D(\hat\theta_{\text{MAP}}))$ & \quad & $\bar D + 2(\bar D-D(\hat\theta_{\text{MAP}}))$ & \quad & $\bar D + \frac{\mathbb{VAR}[D(\theta)|\underline\pi^{-1}]}{2}(\log N -1)$\\
\textbf{DIC}$_\textbf{2}$ & \quad & \textbf{BPIC}$_\textbf{2}$ & \quad & \textbf{BICM}$_\textbf{2}$ \\
\hline
$\bar D + \frac{\mathbb{VAR}[D(\theta)|\underline\pi^{-1}]}{2}$ & \quad & $\bar D + \mathbb{VAR}[D(\theta)|\underline\pi^{-1}]$ & \quad & $D(\hat\theta_{\text{MAP}}) + \frac{\mathbb{VAR}[D(\theta)|\underline\pi^{-1}]}{2}\log N$
\end{tabular}
%\end{center}
%\endsideways
\end{table}
%

%%
%\begin{table}[t]
%\caption{Model selection criteria implemented in the \pkg{PLMIX} package.}
%\small% smaller size
%%\addtolength{\tabcolsep}{-3pt}    % shrinks horizontally
%\renewcommand{\arraystretch}{1.8} % spreads   vertically
%%\sideways
%\label{t:Bayselection}
%%\begin{center}
%\centering
%\begin{tabular}
% {lc} 
%%\multicolumn{2}{c}{Model selection criteria} \\
%Criterion & Formula  \\
%\hline
%DIC$_1$ & $\bar D + (\bar D-D(\hat\theta_{\text{MAP}}))$  \\
%DIC$_2$ & $\bar D + \dfrac{\mathbb{VAR}[D(\theta)|\underline\pi^{-1}]}{2}$ \\
%BPIC$_1$ & $\bar D + 2(\bar D-D(\hat\theta_{\text{MAP}}))$\\ 
%BPIC$_2$ & $\bar D + 2\mathbb{VAR}[D(\theta)|\underline\pi^{-1}]$ \\
%BICM$_1$ & $\bar D + \frac{\mathbb{VAR}[D(\theta)|\underline\pi^{-1}]}{2}(\log N -1)$ \\
%BICM$_2$ & $D(\hat\theta_{\text{MAP}}) + \frac{\mathbb{VAR}[D(\theta)|\underline\pi^{-1}]}{2}\log N$\\
%\hline
%\end{tabular}
%%\end{center}
%%\endsideways
%\end{table}
%%

%For a critical 
%up-to-date review of Bayesian model selection tools, see
%%can be found in
%~\cite{Gelman:Hwang:Vehtari}. 

The \pkg{PLMIX} package includes several Bayesian model selection criteria to compare PL mixture models with a different number of components fitted on the same data set. The considered measures include two alternative versions of each of the following criteria: (i) \textit{Deviance Information Criterion} (DIC), originally defined in~\cite{Spiegelhalter}; (ii) \textit{Bayesian Predictive Information Criterion} (BPIC), proposed by~\cite{Ando} and (iii) \textit{Bayesian Information Criterion-Monte Carlo} (BICM), described in~\cite{Raftery2007}. Their formula
%%
%\begin{itemize}
%\item[-] \textit{Deviance Information Criterion} (DIC); 
%\item[-] \textit{Bayesian Predictive Information Criterion} (BPIC);
%\item[-] \textit{Bayesian Information Criterion-Monte Carlo} (BICM);
%%\item[-] \textit{Widely Applicable Information Criterion} (WAIC);
%%\item[-] \textit{Most Often Best} (MOB) and \textit{Smallest Median Deviance  Criterion} (SMD).
%\end{itemize}
%%
%whose analytic expressions
% of the model choice tools 
are recalled in Table~\ref{t:Bayselection},
where 
%More specifically, 
%%by denoting with 
%let 
$D(\theta)=-2\log L(\theta)$ denotes the \textit{deviance function} and 
%with 
$\bar D=\mathbb{E}[D(\theta)|\underline\pi^{-1}]$ is its posterior expectation. For analytic details, see~\cite{Mollica:Tardella2017}.

As apparent in Table~\ref{t:Bayselection}, we advocate
the use of $\hat\theta_\text{MAP}$ as point estimate for the mixture model parameters, 
instead of the posterior mean $\mathbb{E}[\theta|\underline\pi^{-1}]$, since
%%% . We were motivated by the fact that 
the MAP estimate: (i) straightforwardly provides a meaningful estimate not affected by the LS problem;
(ii) guarantees a positive value of model complexity (\textit{effective number of parameters}); (iii) coincides with the MLE solution $\hat\theta_\text{MLE}$ under uninformative prior specification. 
%following reasons:
 %since, as emphasized by~\cite{Celeux:Forbes}:
% %
%\begin{itemize}
%\item[-] straightforwardly provides a meaningful estimate not affected by the LS problem;
%\item[-] guarantees a positive value of model complexity (\textit{effective number of parameters});
%%$p_{D}$;
%\item[-] coincides with the MLE solution $\hat\theta_\text{MLE}$ under uninformative prior specification. 
%\end{itemize}
%%
It follows that, given the likelihood invariance described in~\ref{sss:LS}, all the considered model comparison measures do not suffer of the presence of LS. Thus, their estimation does not require the preliminary relabeling of the MCMC output.

\subsection{Bayesian model assessment}
\label{ss:ma}
%

%Once the model has been selected, a comprehensive inferential analysis should also contemplate
%%the assessment of its goodness-of-fit, i.e.
%the adequacy of the estimated model in describing the observed data \citep{Gelman:Meng:Stern}.
Evaluating the fitting performance of a parametric model can be less
straightforward in ranking data applications than in other
multivariate 
% context.
contexts. 
%Despite its practical relevance, the fitting performance
%is often neglected by practitioners, especially in the frequentist analysis of ranking data.
% applications.
%One of the main approaches for this issue in
%In the
In the frequentist domain, for example, 
model assessment 
%this issue 
is typically addressed with the computation of 
%this issue is typically addressed with a tail-area probability test, namely
the \textit{p-value} associated to a goodness-of-fit statistic,
such as 
the likelihood ratio 
% ??? the one based on the deviance 
or Pearson's chi-squared test.
%%% test.
%that is the probability that the sampling distribution of the test statistic under the posited model $H$ exceeds the observed value.
%, aimed at evaluating the plausibility of the single sample realization of a test statistic
%sample realization of the statistic
%w.r.t. its sampling distribution under the posited model $H$.
However, in sparse data situations
% characterized by a small
%%, where the 
%sample size $N$ w.r.t. the cardinality $K!$ of 
%%$\mathcal{S_K}$, 
%the ranking space, 
serious issues arise with this approach, since the chi-squared 
%sampling 
distribution of the test statistics under the posited model $H$
%, for example, 
no longer applies.
%%% apply.
%One of the few general purpose instances we could retrieve in the ranking data literature is contained in \cite{Cohen:Mallows}.
\cite{Cohen:Mallows} suggested to overcome this difficulty by comparing observed and expected frequencies regarding relevant partitions of the ranking space. 
%The underlying idea of the goodness-of-fit check is the following: if the model fits adequately the observations, future replications of the data from the posited population are expected to be close, in terms of a suitable discrepancy measure, to the observed sample.
The same approach
% relying on the identification of meaningful groupings of the rankings 
has been successfully applied also within the Bayesian paradigm,
%follow this suggestion with 
%where it is possible to
%%the possibility to
%%the unknown parameters are treated as random entities, allowing the 
%generalize 
where the classical test statistic can be generalized into a parameter-dependent quantity, 
referred to 
%in the literature 
as \textit{discrepancy variable} \citep{Gelman:Meng:Stern,Meng}.
%The resulting advantage is the possibility to compare the observed sample with the entire assumed parametric family averaging w.r.t. the posterior distribution, rather than with the single best fitting solution.
%To our knowledge, the definition of such diagnostic measures in the Bayesian ranked data modeling has not been addressed earlier.
In order to assess the adequacy of the Bayesian PL mixture, the \pkg{PLMIX} package provides diagnostic tools
%develop an appropriate discrepancy variable for parametric ranking models, 
%we concentrate 
derived from two 
%frequently applied 
%popular
significant
summary statistics:
% in ranked data analysis:
\begin{enumerate} %
\item the most-liked item frequency vector
$\underline{r}(\upi^{-1})$, whose generic entry is
$$r_i(\upi^{-1}) =\sum_{s=1}^N I_{[\pi^{-1}_{s}(1)=i]}$$
%counts
%ing 
corresponding to the number of times that item $i$ 
is ranked first;
\item
the PC frequency matrix 
$\tau(\upi^{-1})$,
%=[\tau_{ii'}(\upi^{-1})]$ 
%for $i,i'=1,\dots,K$
whose generic entry is
% in the presence of top partial orderings, the generic entry 
%$\tau_{ii'}$ 
% is given by
%%
%\begin{equation*}
%%\begin{split}
%\tau_{ii'}(\upi^{-1})
%%=\sum_{s=1}^N(1-(1-u_{si})(1-u_{si'}))I_{[\pi_s(i)<\pi_s(i')]}
%=\sum_{s=1}^N(u_{si}+u_{si'}-u_{si}u_{si'})I_{[\pi_s(i)<\pi_s(i')]}
%%\end{split}
%\end{equation*}
%%
%
$$\tau_{ii'}(\upi^{-1})=\sum_{s=1}^N(u_{si}+u_{si'}-u_{si}u_{si'})I_{[\pi_s(i)<\pi_s(i')]}$$
%
%When the sample is entirely composed of complete orderings, the generic entry of $\tau$ is
%%
%\begin{equation}
%\label{e:taufull}
%\tau_{ii'}=\sum_{s=1}^NI_{[\pi_s(i)<\pi_s(i')]}\qquad 1\leq i<i'\leq K,
%\end{equation}
%%
%% $1\leq i<i'\leq K$, 
%counts the number of times that item $i$ 
corresponding to the number of times that item $i$ 
is preferred to item $i'$.
\end{enumerate}
% i.e. the PC matrix $\tau$.
%When the sample is entirely composed of complete orderings, the generic entry of $\tau$ is
%%
%\begin{equation}
%\label{e:taufull}
%\tau_{ii'}=\sum_{s=1}^NI_{[\pi_s(i)<\pi_s(i')]}\qquad 1\leq i<i'\leq K,
%\end{equation}
%%
%% $1\leq i<i'\leq K$, 
%corresponding to the total number of wins of item $i$ over item $i'$.
%Since ties are not contemplated in this setting, the sum $T_{ii'}=\tau_{ii'}+\tau_{i'i}$ indicating the total number of pairwise comparisons between item $i$ and $i'$ is equal to the sample size $N$.
%In the case of partial orderings, definition \eqref{e:taufull} becomes
%%the generic entry in $\tau$ is given by
%%
%\begin{equation*}
%\begin{split}
%\tau_{ii'}=\sum_{s=1}^N(1-(1-u_{si})(1-u_{si'}))I_{[\pi_s(i)<\pi_s(i')]}
%=\sum_{s=1}^N(u_{si}+u_{si'}-u_{si}u_{si'})I_{[\pi_s(i)<\pi_s(i')]}.
%\end{split}
%\end{equation*}
%%
%%Denoting with $T_{ii'}$ the total number of pairwise comparisons between item $i$ and $i'$, one has $$T_{ii'}=\tau_{ii'}+\tau_{i'i}=\gamma_i+\gamma_{i'}-\sum_{s=1}^Nu_{si}u_{si'},$$
%%which coincides with the number of sample units who ranked at least one of the two items. Obviously, if the sample is entirely composed of complete orderings, $T_{ii'}$ reduces to the sample size $N$.
%As described in \cite{Mollica:Tardella2017}, 
One could then employ the two
%aforementioned 
sample quantities to define the 
chi-squared discrepancies $X^2_{(1)}(\upi^{-1};\theta)$ and $X^2_{(2)}(\upi^{-1};\theta)$ comparing observed and expected frequencies under the PL mixture scenario $H$, see~\cite{Mollica:Tardella2017} for the explicit formulas. 
For a given discrepancy variable $X^2(\upi^{-1};\theta)$,
%perform a 
the posterior predictive check of model goodness-of-fit relies on the computation of the
%(see \cite{Gelman:Meng:Stern} for illustrative examples)
%We propose the following PCs-based test
%%
%\begin{equation*}
%\label{e:discr}
%X^2(\upi_{\text{rep}}^{-1 (l)};\theta^{(l)})\geq X^2(\upi^{-1}_\text{obs};\theta^{(l)})
%=\sum_{i<i'}\frac{\left(\tau_{ii'}^{(l)}
%-T_{ii'}^{(l)}\frac{p_i^{(l)}}{p_i^{(l)}+p_{i'}^{(l)}}\right)^2}{T_{ii'}^{(l)}\frac{p_i^{(l)}}{p_i^{(l)}+p_{i'}^{(l)}}}
%\geq \sum_{i<i'}\frac{\left(\tau_{ii'}
%-T_{ii'}\frac{p_i^{(l)}}{p_i^{(l)}+p_{i'}^{(l)}}\right)^2}{T_{ii'}\frac{p_i^{(l)}}{p_i^{(l)}+p_{i'}^{(l)}}}.
%\end{equation*}
%%
%We propose to employ t
%The 
\textit{posterior predictive $p$-value} 
%$p_B$, to be employed as Bayesian fitting measure
%in a posterior predictive check
%for checking the model fitness is then obtained 
%corresponding to
%given by
%
\begin{equation}
\label{e:postp}
p_B=\PP(X^2(\upi_{\text{rep}}^{-1};\theta)\geq X^2(\upi^{-1}_\text{obs};\theta)|\upi^{-1}_\text{obs},H),
\end{equation}
%
%where the randomness is induced by the joint posterior distribution of $(\upi_\text{rep}^{-1},\theta)$ given $H$.
%As emphasized by \cite{Gelman:Meng:Stern}, 
that can be easily approximated
once an MCMC sample
% of size $L$ 
from the posterior distribution is available \citep{Gelman:Meng:Stern}.
% the estimation of \eqref{e:postp} 
%%does not require additional effort 
%can be obtained by iterating 
%%of
%%, but the application of 
%the following two-step procedure for all the parameter draws $\theta^{(l)}$:
%(i) simulate a new data set $\upi_{\text{rep}}^{-1 (l)}$ from the sampling distribution induced by $\theta^{(l)}$; (ii) 
%%subsequently 
%compare $X^2(\upi_{\text{rep}}^{-1 (l)};\theta^{(l)})$ with $X^2(\upi^{-1}_{\text{obs}};\theta^{(l)})$. The posterior predictive $p$-value~\eqref{e:postp} is finally estimated as the 
%%empirical relative frequency 
%proportion of the times that $X^2(\upi_{\text{rep}}^{-1 (l)};\theta^{(l)})$ exceeds $X^2(\upi^{-1}_{\text{obs}};\theta^{(l)})$. 
%Graphically, we can visualize the Bayesian fitting measure \eqref{e:postp} in the $(X^2_{\text{obs}},X^2_{\text{rep}})$ plane as the proportion of points above the 45$^\circ$ line.
%%From follows that the $p_B$ can be interpreted as the expectation of the classical p-value w.r.t. the posterior distribution of the parameter vector. Clearly, when the chosen discrepancy is not parameter-dependent under the assumed population, the two criteria coincide.
%Given the occurrence of partial orderings in our applications, we simply reproduced the same missingness pattern $\{n_s\}$ observed in the sample on the drawings $\upi_{\text{rep}}^{-1 (l)}$ from the posterior predictive distribution.
Clearly, an efficient simulation device is needed to assist the drawing of replicated datasets $\upi_{\text{rep}}^{-1}$ from the posterior predictive distribution.  

\cite{Mollica:Tardella2017} also showed the usefulness of model assessment conditionally on the number  $m=1,\dots,K-1$ of ranked items.
%, in order to better highlight possible deficiencies of the fitted model to reproduce specific features of the observed data.
% within each subsample of partial sequences with the same length. 
To this aim, they introduced two additional discrepancies $\tilde X^2_{(1)}$ and $\tilde X^2_{(2)}$ that
parallel $X^2_{(1)}$ and $X^2_{(2)}$, 
%%% defines as
%since they are defined as
given by
%%
%{\small
%\begin{equation}
%\label{e:chitilde}
%\tilde X^2_{(1)}(\upi^{-1};\theta)=\sum_{m=1}^{K-1}\sum_{i=1}^K\frac{(r_{i,m}-r^*_{i,m}(\theta))^2}{r^*_{i,m}(\theta)}
%%\quad\text{and}\quad 
%\qquad\qquad
%\tilde X^2_{(2)}(\upi^{-1};\theta)= \sum_{m=1}^{K-1}\sum_{i<i'}\frac{(\tau_{ii',m}-\tau^*_{ii',m}(\theta))^2}{\tau^*_{ii',m}(\theta)},
%\end{equation}}
%%
%
\begin{equation*}
\label{e:chitilde}
\tilde X^2_{(1)}(\upi^{-1};\theta)=\sum_{m=1}^{K-1}X^2_{(1)}(\upi^{-1}_m;\theta)
%\quad\text{and}\quad 
\qquad\qquad
\tilde X^2_{(2)}(\upi^{-1};\theta)= \sum_{m=1}^{K-1}X^2_{(2)}(\upi^{-1}_m;\theta),
\end{equation*}
where the presence of $m$ in the subscript refers to the evaluation of
%$r_i$ and $\tau_{ii'}$ 
the discrepancies in the subsample $\upi^{-1}_m=\{\pi^{-1}_s:
n_s=m\}$. The computation of $\tilde p_B$, obtained
from equality~\eqref{e:postp} by replacing $X^2$ with
% the generic discrepancy 
$\tilde X^2$,
permits to assess the adequacy of the model estimated 
%%% on 
on the entire dataset to recover the considered summary statistics within the subsets of partial orderings with the same length.

Finally, similarly to the model comparison step, the LS adjustment of the posterior samples is not necessary for the posterior predictive check. This is due to
%can be explained with 
the use of the marginal support parameters $p_i=\sum_{g=1}^G\omega_gp_{gi}$ in the computation of the expected frequencies,
% involved in the chi-squared discrepancies, 
which are invariant to the LS phenomenon.
% but, once the adequacy of the selected model has been assessed, for the computation of the point estimates from the corresponding posterior MCMC sample. 

\section{The PLMIX package}
\label{s:package}
\begin{table}[t]
\centering
\begin{threeparttable}
\caption{Classification of the 23 objects included in the novel \pkg{PLMIX} package.}
\addtolength{\tabcolsep}{-3pt}
\renewcommand{\arraystretch}{1.2} % spreads   vertically
\begin{tabular}{ccccc}
\multicolumn{2}{c}{\textbf{Ranking data manipulation}}  & &  \multicolumn{2}{c}{\textbf{Ranking data simulation}}  \\ 
\cline{1-2}\cline{4-5}
%\rowcolor{LightCyan}
%\rowcolor{Gray}
\color{cyan}{\code{binary_group_ind}} & \color{cyan}{\code{freq_to_unit}} & & \multicolumn{2}{c}{\color{cyan}{\code{rPLMIX}}} \\ 
\color{cyan}{\code{make_complete}} &  \color{cyan}{\code{make_partial}} & & \multicolumn{2}{c}{\textbf{Ranking data description}}  \\ 
\cline{4-5}
\color{cyan}{\code{rank_ord_switch}} & \color{cyan}{\code{unit_to_freq}} & &
\color{cyan}{ \code{paired_comparisons}} & \color{cyan}{\code{rank_summaries}}  \\ 
\multicolumn{2}{c}{\textbf{Model estimation}} & & \multicolumn{2}{c}{\textbf{Model selection}}   \\ 
\cline{1-2}\cline{4-5}
\color{cyan}{ \code{likPLMIX}} & \color{cyan}{\code{loglikPLMIX}} & & \color{cyan}{ \code{selectPLMIX}} & \color{cyan}{\code{bicPLMIX}} \\ 
\color{cyan}{ \code{mapPLMIX}} & \color{cyan}{\code{mapPLMIX_multistart}} & & \multicolumn{2}{c}{\textbf{Model assessment \& \textbf{LS}}} \\ 
\cline{4-5}
\color{cyan}{\code{gibbsPLMIX}} & & & \color{cyan}{\code{ppcheckPLMIX}} & \color{cyan}{\code{ppcheckPLMIX_cond}} \\ 
 & & & \color{cyan}{\code{label_switchPLMIX}} &  \\ 
 \multicolumn{5}{c}{\textbf{Data}} \\
\hline
\multicolumn{5}{c}{\color{BurntOrange}{\code{d\char`_apa}}\qquad\color{BurntOrange}{\code{d\char`_carconf}}\qquad\color{BurntOrange}{\code{d\char`_dublinwest}}\qquad\color{BurntOrange}{\code{d\char`_german}}\qquad\color{BurntOrange}{\code{d\char`_nascar}}}  \\
\hline
\end{tabular}
\label{t:overview}
\begin{tablenotes}
      \small
%      \item ISR $=$ Insertion Sort Rank model, GMM $=$ Generalized Mallows model, MNOS $=$ multivariate normal ordered statistic model, WDB $=$ weighted distance-based model  
      \item[\color{cyan}{$\blacksquare$}] object of class \code{"function"}\qquad\color{BurntOrange}{$\blacksquare$} \color{black}{object of class \code{"matrix"}}
\end{tablenotes}
\end{threeparttable}
\end{table}

The novel \pkg{PLMIX} is 
%%% entirely 
%It is 
the first \proglang{R} package
% specifically
devoted to Bayesian inference for partially ranked data. 
More specifically, \pkg{PLMIX} performs Bayesian estimation of ranking models
% in \proglang{R} 
by focusing on the PL and its finite mixture extension as the sampling distribution. 
In the present setting, 
%as better detailed in Section~\ref{sss:MAPhete}, 
the MLE approach is recovered as a special case of the Bayesian analysis 
with a noninformative (flat) prior specification.
%in the frequentist domain. 

To address the issue of computationally demanding procedures, typical in ranking contexts,
%related to the inference on the PL parameters 
\pkg{PLMIX} can take advantage of a hybrid code linking
%by combining 
the \proglang{R} environment with the \proglang{C++} programming language. The parallelization option is also implemented, such that finite mixtures with a different number of components can be simultaneously analyzed.
% by the user.
% To further reduce the computational time.

%The 
\pkg{PLMIX} 
%package 
contains 24 objects visible to the user, classified 
%provides an overview of the available objects classified 
according the their task in Table~\ref{t:overview}. There are 19 objects of class \code{"function"} and 5 datasets.
%10 are functions 
As revealed by the overview,
% in Table~\ref{t:overview}, 
the novel 
%\pkg{PLMIX}
package provides a suite of functions assisting each step of the ranking data
analysis.
In fact, 
in addition to data manipulation tools, descriptive summary and estimation techniques, the package assists other fundamental phases related to the PL mixture analysis, such that the selection of the optimal number of components and the goodness-of-fit assessment, aimed at a more critical exploration of the group structure in the sample.
%
%the available functions accomplish ranking
%data manipulation, descriptive summarization, simulation, model
%estimation procedures, model selection criteria and model assessment
%diagnostics. 
Also the treatment of the LS problem is supported in 
our package.
%\code{bicPLMIX}, \code{gibbsPLMIX}, \code{likPLMIX}, \code{loglikPLMIX}, \code{mapPLMIX}, \code{mapPLMIX_multistart}, \code{ppcheckPLMIX}, \code{ppcheckPLMIX_cond}, \code{rPLMIX}, \code{selectPLMIX}
%needed for the simulation and inferential process for
%fitting and selecting the appropriate PLMIX model.
%The remaining 8 functions
%\code{binary_group_ind},  \code{freq_to_unit},  \code{make_complete},  \code{make_partial},  \code{paired_comparisons},  \code{rank_ord_switch},  \code{rank_summaries},  \code{unit_to_freq}
%are mainly for ranking data manipulation and summarization.
The 5 datasets are all provided in ordering format as objects of class \code{"matrix"}. 
%%Since for top-$t$ partial orderings only the first $t$ positions are assigned, the remaining missing $K-t$ ones have to be 
Missing positions/items in the partial top orderings are denoted with zero entries and Rank $=$ 1 indicates the most-liked alternative. 
%(\code{d_apa}, \code{d_carconf}, \code{d_dublinwest}, \code{d_german}, \code{d_nascar})

The next subsections illustrate in greater detail the application of the \pkg{PLMIX} commands to simulated and real ranking data.

\subsection{Ranking data manipulation: Dublin West and German sample data}
\label{ss:data}
Before performing a ranking data analysis, it is important to know
exactly the data format
% of the data 
and employ the suitable one, in order
to avoid 
erroneous implementation or 
misleading interpretations.
% of the results.
%The same observed sequence, in fact, can lead to two different interpretations.
%As stressed in Section~\ref{ss:data}, the observed ordered $K$-tuple
%can be read in two different ways. Here is a toy example with $K=4$
%items, where the vector (3,1,4,2) could be regarded either as a
%complete {\em ranking} or a complete {\em ordering} but with a substantial different interpretation, that is,
%%Here is a simple example with $K=4$ items
%$$\pi=(3,1,4,2)\qquad\neq\qquad\pi^{-1}=(3,1,4,2);$$
%in fact, in the former notation $\pi$
%%the same vector $(3,1,4,2)$ indicates that
%the first position is occupied by Item 2 (since $\pi(2)=1$),
%% if assumed as a ranking $\pi$,
%whereas in the latter interpretation $\pi^{-1}$ the top position is occupied by Item 3 (since $\pi^{-1}(1)=3$).
% if interpreted as an ordering $\pi^{-1}$ .
%In order to avoid confusion, we will henceforth make explicit use of the inverse function notation when we refer to orderings.
%Consequently, 
The preliminary conversion of the data into the appropriate format
% is needed to start the ranking data analysis.
%when the original data set is expressed in the form of ranking. 
%This transformation 
can be performed by means of the
\code{rank_ord_switch} function,
% of the \pkg{PLMIX} package, 
switching
from orderings to rankings and vice-versa for both complete and partial
observations. 
%For partial sequences (ranking/ordering), the missing positions/items have to be denoted with zero entries. For example, $\pi=(0,1,0,2)$ corresponds to a top-2 ranking and can be equivalently expressed as top-2 ordering with $\pi^{-1}=(2,4,0,0)$. 
The following instructions show the simple application of the \code{rank_ord_switch} routine to the first 6 partial orderings of the 2002 Dublin West election dataset \citep{Mattei:Walsh:2013} called \code{d_dublinwest}, in order to convert them into the ranking format. After loading the package and the data
\begin{CodeChunk}
\begin{CodeInput}
> data(d_dublinwest)
> head(d_dublinwest)
\end{CodeInput}
\begin{CodeOutput}
     rank1 rank2 rank3 rank4 rank5 rank6 rank7 rank8 rank9
[1,]     7     9     4     2     8     0     0     0     0
[2,]     5     3     7     6     0     0     0     0     0
[3,]     5     7     3     0     0     0     0     0     0
[4,]     9     2     7     0     0     0     0     0     0
[5,]     3     2     0     0     0     0     0     0     0
[6,]     5     3     2     0     0     0     0     0     0
\end{CodeOutput}
\end{CodeChunk}
one can apply the command as follows
\begin{CodeChunk}
\begin{CodeInput}
> rank_ord_switch(data=head(d_dublinwest), format="ordering")
\end{CodeInput}
\begin{CodeOutput}
     [,1] [,2] [,3] [,4] [,5] [,6] [,7] [,8] [,9]
[1,]    0    4    0    3    0    0    1    5    2
[2,]    0    0    2    0    1    4    3    0    0
[3,]    0    0    3    0    1    0    2    0    0
[4,]    0    2    0    0    0    0    3    0    1
[5,]    0    2    1    0    0    0    0    0    0
[6,]    0    3    2    0    1    0    0    0    0
\end{CodeOutput}
\end{CodeChunk}
where the input arguments are: i) \code{data}: the numeric $N\times K$ data
matrix of partial sequences to be converted, ii) \code{format}: the character string
indicating the format of the input \code{data}
% argument 
and iii) \code{nranked}: the optional numeric vector of length $N$ with the number of items ranked by each sample unit (default is \code{NULL}).

Another useful task is the aggregation of the replicated sequences in the observed dataset. The \code{unit_to_freq} routine constructs the frequency distribution of the observed sequences from the dataset of individual rankings/orderings supplied in the single argument \code{data}. Here is the output of \code{unit_to_freq} when applied to the German Sample dataset \code{d_german}, collecting complete orderings of $K=4$ political goals
\begin{CodeChunk}
\begin{CodeInput}
> data(d_german)
> unit_to_freq(data=d_german)
\end{CodeInput}
\begin{CodeOutput}
      [,1] [,2] [,3] [,4] [,5]
 [1,]    1    2    3    4  137
 [2,]    1    2    4    3   29
 [3,]    1    3    2    4  309
 [4,]    1    3    4    2   52
 [5,]    1    4    2    3  255
 [6,]    1    4    3    2   93
 [7,]    2    1    3    4   48
 [8,]    2    1    4    3   23
 [9,]    2    3    1    4  330
[10,]    2    3    4    1   21
[11,]    2    4    1    3  294
[12,]    2    4    3    1   30
[13,]    3    1    2    4   61
[14,]    3    1    4    2   33
[15,]    3    2    1    4  117
[16,]    3    2    4    1   29
[17,]    3    4    1    2   70
[18,]    3    4    2    1   35
[19,]    4    1    2    3   55
[20,]    4    1    3    2   59
[21,]    4    2    1    3   69
[22,]    4    2    3    1   52
[23,]    4    3    1    2   34
[24,]    4    3    2    1   27
\end{CodeOutput}
\end{CodeChunk}
The observed frequencies are indicated in the last $(K+1)$-th column. The frequency distribution helps to explore the possible presence of multimodal patterns in the sample and to compare the observed frequencies with those expected under specific parametric assumptions. Additionally, it can be exploited to prepare the data for the analysis with methods implemented in other \proglang{R} packages requiring the aggregate format. 

Conversely, the \code{freq_to_unit} function expands the frequency distribution supplied in the argument \code{freq_distr}
% of the observed sequences 
into the dataset of individual rankings/orderings.
%, as described 
In the following toy example, we consider a synthetic sample of size $N=6$ with 2 top-1, 1 top-2 and 3 top-3 partial rankings
\begin{CodeChunk}
\begin{CodeInput}
> obs_rankings <- rbind(c(0,0,1,0), c(0,1,0,2), c(4,1,2,3))
> freq_to_unit(freq_distr=cbind(obs_rankings, c(2,1,3)))
\end{CodeInput}
\begin{CodeOutput}
     [,1] [,2] [,3] [,4]
[1,]    0    0    1    0
[2,]    0    0    1    0
[3,]    0    1    0    2
[4,]    4    1    2    3
[5,]    4    1    2    3
[6,]    4    1    2    3
\end{CodeOutput}
\end{CodeChunk}

Further helpful commands for
% a preliminary 
data manipulation are \code{make_partial} and
\code{make_complete},
% commands, 
that can be regarded 
specularly.
%%% as two opposites. 
The former allows for the truncation of complete sequences according
to different censoring patterns, either in a 
%%% determinist
deterministic 
or a random way. 
The deterministic approach requires the user to specify the number of top positions to be retained for each sample unit in the \code{nranked} argument. The random approach, instead, makes use of the probabilities of top-1, top-2, \dots, top-$(K-1)$ censoring patterns, supplied in the \code{probcens} vector, to perform a stochastic truncation of the complete sequences. 
%Recall that a top-(K-1) sequence corresponds to a complete ordering\ranking. The returned \code{partialdata} matrix has the same format of the input \code{data} with missing positions denoted with zero entries.
For example,
% with the following instructions one can apply 
a random truncation
% to the complete orderings 
of the \code{d_german} dataset with a 60\% overall rate of censored observations and equal chance of top-1 and top-2 orderings can be obtained with the following code
\begin{CodeChunk}
%\begin{CodeInput}
%%> data(d_german)
%> head(d_german)
%\end{CodeInput}
%\begin{CodeOutput}
%     [,1] [,2] [,3] [,4]
%[1,]    1    2    3    4
%[2,]    1    2    3    4
%[3,]    1    2    3    4
%[4,]    1    2    3    4
%[5,]    1    2    3    4
%[6,]    1    2    3    4
%
%\end{CodeOutput}
\begin{CodeInput}
> set.seed(57524)
> d_german_cens <- make_partial(data=d_german, format="ordering", 
+   probcens=c(0.3, 0.3, 0.4))
\end{CodeInput}
\end{CodeChunk}
It returns a list with two named objects given by the numeric data matrix \code{partialdata} of censored sequences and the numeric vector \code{nranked} with the number of items ranked by each sample unit after the random censoring. Here is the code to extract them and to verify the consistency of the resulting censored dataset with the nominal probability values specified in the \code{probcens} argument
\begin{CodeChunk}
\begin{CodeInput}
> head(d_german_cens$partialdata)
\end{CodeInput}
\begin{CodeOutput}
     [,1] [,2] [,3] [,4]
[1,]    1    2    0    0
[2,]    1    0    0    0
[3,]    1    2    0    0
[4,]    1    2    3    4
[5,]    1    2    0    0
[6,]    1    0    0    0
\end{CodeOutput}
\begin{CodeInput}
> round(table(d_german_cens$nranked)/nrow(d_german), 2)
\end{CodeInput}
\begin{CodeOutput}
   1    2    4 
0.30 0.29 0.41 
\end{CodeOutput}
\end{CodeChunk}
The \code{make_partial} function is especially useful in simulation studies to investigate the impact of the censoring mechanism on the ability of the estimation procedures to recover the true generating distribution and, additionally, to verify their robustness to the censoring rate. See, for example, the simulation study in~\cite{Mollica:Tardella2017}.

Conversely, the \code{make_complete} function is conceived for the completion of partial orderings by filling in the missing (zero) positions/items 
%of the partial sequences 
with the remaining not-selected alternatives. More specifically, the
completion of the partial data is performed with the random procedure
determined the Plackett-Luce scheme, that is, with a sampling without
replacement of the 
%not-ranked 
unranked items. To this aim, the positive values specified in the \code{probitems} argument are used as support parameters. For instance, the random completion of the \code{d_dublinwest} dataset with decreasing support over the $K=9$ candidates can be implemented as follows
\begin{CodeChunk}
%\begin{CodeInput}
%%> data(d_dublinwest)
%> head(d_dublinwest)
%\end{CodeInput}
%\begin{CodeOutput}
%     rank1 rank2 rank3 rank4 rank5 rank6 rank7 rank8 rank9
%[1,]     7     9     4     2     8     0     0     0     0
%[2,]     5     3     7     6     0     0     0     0     0
%[3,]     5     7     3     0     0     0     0     0     0
%[4,]     9     2     7     0     0     0     0     0     0
%[5,]     3     2     0     0     0     0     0     0     0
%[6,]     5     3     2     0     0     0     0     0     0
%\end{CodeOutput}
%\begin{CodeInput}
%> top_freq <- rank_summaries(data=d_dublinwest, format="ordering", 
%+   mean_rank=FALSE, pairedcomparisons=FALSE)$marginals["Rank 1",]
%> top_freq
%\end{CodeInput}
%\begin{CodeOutput}
%[1]  748 3810 2300 6442 8086 2404 2370  134 3694
%\end{CodeOutput}
\begin{CodeInput}
%> K <- ncol(d_dublinwest)
> set.seed(57524)
> d_dublinwest_compl <- make_complete(data=d_dublinwest, format="ordering", 
+   probitems=ncol(d_dublinwest):1)
> head(d_dublinwest_full$completedata)
\end{CodeInput}
\begin{CodeOutput}
     rank1 rank2 rank3 rank4 rank5 rank6 rank7 rank8 rank9
[1,]     7     9     4     2     8     3     6     5     1
[2,]     5     3     7     6     2     8     1     4     9
[3,]     5     7     3     4     1     2     8     6     9
[4,]     9     2     7     1     4     3     6     5     8
[5,]     3     2     5     6     4     7     1     8     9
[6,]     5     3     2     1     7     8     4     6     9
\end{CodeOutput}
\end{CodeChunk}
Other possible input values for the vector \code{probitems} could be the observed frequencies that each item has been ranked in the first position, in order to preserve the univariate feature %regarding the top preferences
of the observed sample.
%Finally, we also remind that a top-$(K-1)$ sequence
%%In this regard, that the case $t=K-1$ 
%corresponds to a complete observation.
%%% , since the single missing entry is unambiguously determined.  
%To optimize the computational time, the preliminary specification of
%the bottom position in top-$(K-1)$ sequences is recommended.
%%For example, we could explicit the last entry of partial orderings of length $K-1$, because it is  unambiguously determined, as follows (METTI ESEMPIO)

\subsection{Ranking data simulation and likelihood function: simulated data}
\label{ss:simulation}
Data simulation and likelihood function are 
% one of the prime tasks to be addressed 
essential tasks to be suitably implemented in 
%%% by 
a model-oriented statistical package. 
The random generation of complete orderings is accomplished with the \code{rPLMIX} routine. 
A random sample of $N=5$ complete orderings of $K=6$ items can be drawn from a $3$-component PL mixture with parameters
$$\underline{p}=  
\begin{pmatrix}
    1 & 2 & 3 & 4 & 5 & 6 \\
    6 & 5 & 4 & 3 & 2 & 1  \\
    1 & 1 & 1 & 1 & 1 & 1 \\
  \end{pmatrix}
\qquad\uomega=(0.50, 0.25, 0.25)$$
with the following instructions
\begin{CodeChunk}
\begin{CodeInput}
> K <- 6
> p_par <- rbind(1:K, K:1, rep(1, K))
> w_par <- c(0.50, 0.25, 0.25)
> set.seed(57524)
> simulated_data <- rPLMIX(n=5, K=K, G=3, p=p_par, weights=w_par,
+   format="ordering")
\end{CodeInput}
\end{CodeChunk}
where the argument \code{p} requires the numeric $G\times K$ matrix of the component-specific support parameters
% in the rows 
and \code{weights} is the vector of mixture weights.
If $G>1$, the \code{rPLMIX} function returns a list of two
named objects corresponding, respectively, to the
%%% 
vector \code{comp} of simulated
component memberships and to the 
%%% 
matrix \code{sim_data} of 
simulated orderings, given by
\begin{CodeChunk}
\begin{CodeInput}
sim_orderings$comp
\end{CodeInput}
\begin{CodeOutput}
[1] 1 2 3 1 1
\end{CodeOutput}
\begin{CodeInput}
sim_orderings$sim_data
\end{CodeInput}
\begin{CodeOutput}
     [,1] [,2] [,3] [,4] [,5] [,6]
[1,]    2    6    5    4    1    3
[2,]    2    4    3    1    5    6
[3,]    3    6    2    1    5    4
[4,]    4    3    5    6    1    2
[5,]    3    2    5    6    4    1
\end{CodeOutput}
\end{CodeChunk}

As evident in equation~\eqref{e:plobsloglik}, the calculation of the PL log-likelihood is computationally intensive, especially for large data sets, since the normalization of the support parameters varies across sample units and is 
performed sequentially
%needed at each stage of the sequential 
in the ranking process. Of course, the computational demand increases in the finite mixture setting. On the other hand, an efficient evaluation of the likelihood is crucial for the application of iterative optimization methods such as the EM algorithm, both in the MLE perspective and in the MAP estimation detailed in Section~\ref{sss:MAPhete}. 
%Likelihood evaluation is also required for the computation of Bayesian model comparison criteria. 
In this regard, the \code{loglikPLMIX} function included in the \pkg{PLMIX} package 
%for the computation of~\eqref{e:plobsloglik} 
calls a \proglang{C++} routine from \proglang{R}
% environment 
to reduce the computational 
%%% time. 
burden. To show 
%The following instructions 
%%%% compares 
%compare
%the computational time of 
the efficiency of the \code{loglikPLMIX} function 
for the
evaluation of the log-likelihood~\eqref{e:plobsloglik}, we first simulated a large
dataset of $N=15000$ orderings of $K=6$ items
% has been 
%%%% generate 
%generated 
%%% 
from the (default) uniform ranking model, corresponding to the PL with
%%the special case of~\eqref{e:pl} with 
constant support parameters
% $p_i=p$ for all $i=1,\dots,K$ 
% in terms of the computational times; 
%The example considers the log-likelihood evaluation for the $5738$ full rankings from the popular American Psychological Association (APA) election data set, available in the \pkg{Rankcluster} package, under the uniform ranking model, the special case of~\eqref{e:pl} with constant support parameters $p_i=p$ for all $i=1,\dots,K$:
%
\begin{CodeChunk}
\begin{CodeInput}
> K <- 6
> set.seed(57524)
> unif_data <- rPLMIX(n=15000, K=K, G=1, format="ordering")
\end{CodeInput}
\end{CodeChunk}
Then we have compared 
the time needed to obtain the maximized log-likelihood value with \code{loglikPLMIX} and with the
\code{Likelihood.PL} command of the \pkg{StatRank} package 
\begin{CodeChunk}
\begin{CodeInput}
> PLpar <- rep(1, K)
> system.time(loglikPLMIX(p=t(PLpar), ref_order=t(1:K), weights=1,
+   pi_inv=unif_data))
\end{CodeInput}
\begin{CodeOutput}
   user  system elapsed 
  0.005   0.000   0.005 
\end{CodeOutput}
\begin{CodeInput}
> library(StatRank)
> system.time(Likelihood.PL(Data=unif_data, parameter=list(m=K, Mean=PLpar)))
\end{CodeInput}
\begin{CodeOutput}
   user  system elapsed 
  0.181   0.002   0.182 
\end{CodeOutput}
\end{CodeChunk}
%
%As evident, the \code{loglikPLMIX} function significantly outperforms the alternative \code{Likelihood.PL} command. 

Finally, notice that the \code{rPLMIX} and \code{loglikPLMIX} functions share the
\code{ref_order} argument relative to the \textit{reference order}
parameters of the mixture of Extended Plackett-Luce models (EPL)
introduced by~\cite{Mollica:Tardella}. The traditional PL is a special
instance of the EPL with reference order parameter equal to the identity
permutation $(1,\dots,K)$. Since the current version of \pkg{PLMIX}
implements the mixture of PL models, the \code{ref_order} argument
must be a matrix with 
%%% the 
$G$ rows equal to the identity permutation.

\subsection{Ranking data description: CARCONF data}
\label{ss:descriptive}
Useful utilities to conduct a preliminary exploratory analysis
% before fitting a ranking model 
are included in \pkg{PLMIX}. Unlike similar functions from other packages, these functions can handle partial observations. To this purpose,
%For descriptive purposes, 
the main command is named
\code{rank_summaries} that 
accomplishes
%combines several sample statistics. In particular, it assists 
the computation of summary statistics and censoring patterns for a partial ordering/ranking dataset. 
The basic application of the  \code{rank_summaries} routine requires the same inputs (\code{data} and \code{format})
% are defined as in the 
of the \code{rank_ord_switch} function. For the \code{d_carconf} dataset, the command returns the following information 
\begin{CodeChunk}
\begin{CodeInput}
> data(d_carconf)
> rank_summaries(data=d_carconf, format="ordering")
\end{CodeInput}
\begin{CodeOutput}
$nranked
  [1] 6 6 6 6 4 4 3 6 6 6 3 6 6 4 6 2 6 6 6 6 6 6 2 6 6 6 6 6 6 6 6 6 6
 [34] 3 6 6 6 6 6 6 6 6 6 3 6 6 6 6 6 4 6 4 6 3 3 4 6 6 6 3 6 4 6 6 6 6
 [67] 6 6 6 6 6 6 6 6 6 6 6 6 6 6 6 6 6 6 6 6 6 4 6 6 6 6 6 6 6 3 3 6 6
[100] 6 4 4 6 3 4 4 6 6 6 6 6 6 4 6 6 6 6 4 6 6 4 6 6 4 6 6 6 6 6 6 6 6
[133] 6 6 6 6 6 6 4 4 4 6 6 6 6 4 6 6 6 6 6 6 6 6 6 6 6 6 6 6 6 4 4 6 3
[166] 6 6 6 6 6 6 6 3 6 6 3 6 6 6 6 6 6 6 4 6 6 6 4 6 6 4 6 6 6 6 6 6 6
[199] 6 6 3 3 6 6 6 6 6 6 6 2 6 6 2 6 6 4 6 6 6 6 2 6 6 6 6 6 6 6 6 6 6
[232] 6 4 6 6 6 6 6 6 6 6 2 6 6 6 6 6 6 6 6 6 6 6 6 6 6 6 4 6 6 4 6 3 4
[265] 4 6 6 6 6 6 6 6 6 6 2 6 6 6 6 6 6 6 6 6 6 6 6 6 6 6 6 6 6 6 6 4 6
[298] 6 6 6 1 6 6 6 6 6 6 6 6 6 6 6 6 6 4 6 6 6 6 6 6 6 6 4 6 2 6 6 6 4
[331] 6 6 6 4 6 6 6 6 6 6 6 6 6 6 6 6 6 6 6 6 6 6 6 6 6 6 6 6 6 6 6 6 6
[364] 6 6 6 6 6 6 6 6 6 4 6 6 4 6 6 6 6 6 6 6 6 6 6 6 6 6 6 6 6 6 6 6 6
[397] 6 6 6 6 6 6 6 6 4 4 6 6 6 6 6 3 4 3 6 6 6 6 6 6 6 6 6 6 6 6 6 6 6
[430] 6 6 4 6 4 6

$nranked_distr
Top-1 Top-2 Top-3 Top-4 Top-6 
    1     8    18    43   365 

$missing_pos
[1] 42 17  0 29 62 27

$mean_rank
[1] 3.559796 2.882775 3.165517 3.113300 4.493298 3.203431

$marginal_rank_distr
       Item 1 Item 2 Item 3 Item 4 Item 5 Item 6
Rank 1     86    101     87     78     28     55
Rank 2     53     87     85     86     27     96
Rank 3     46     84     78     76     46     96
Rank 4     61     74     81     69     51     72
Rank 5     57     50     62     72     74     50
Rank 6     90     22     42     25    147     39

$pairedcomparisons
       Item 1 Item 2 Item 3 Item 4 Item 5 Item 6
Item 1      0    171    179    178    250    181
Item 2    257      0    238    237    329    242
Item 3    256    197      0    230    324    226
Item 4    243    193    205      0    306    225
Item 5    148     92    111    105      0    106
Item 6    239    187    209    199    307      0
\end{CodeOutput}
\end{CodeChunk}
%
%where the first element of the list has been omitted for reasons of space. 
The resulting list includes the following named objects:
\begin{description}[leftmargin=!,labelwidth=\widthof{\code{missing\char`_positions}}]
\item[\code{nranked}] numeric vector with the number of items ranked by each sample unit;
\item[\code{nranked\char`_distr}] the frequency distribution of the \code{nranked} vector;
\item[\code{missing\char`_positions}] numeric vector with the number of missing positions for each item;
\item[\code{mean\char`_rank}] numeric vector with the mean rank of each item;
\item[\code{marginals}] numeric $K\times K$ matrix of the marginal rank distributions;
\item[\code{pairedcomparisons}] numeric $K\times K$ matrix of PCs.
\end{description}
Specifically, the first row of the matrix \code{marginals}, labeled as \code{Rank 1}, corresponds to the vector $\underline{r}(\upi^{-1})$, whereas the matrix \code{pairedcomparisons} is $\tau(\upi^{-1})$. 
%Finally, as 
The command \code{rank_summaries} has additional logical arguments
% \code{mean_rank}, \code{marginals} and \code{pairedcomparisons}, 
indicating, respectively, whether the mean rank vector, the marginal rank distribution and the PC frequencies have to be actually computed (default is \code{TRUE}). The PC matrix is implemented in \proglang{C++} to speed up the execution and can be separately computed also with the \code{paired_comparisons} function. 

As better detailed in Section~\ref{ss:assess}, descriptive summaries are also involved in the model assessment step to investigate the compatibility between the observed dataset and specific parametric assumptions. Thus, their efficient implementation is crucial to reduce the computational time needed for the goodness-of-fit diagnostics.
%, to check the adequacy of the estimated model to recover some features of the observed data.

\subsection{Model estimation: APA data}
\label{ss:estimate}

The core 
%%% 
inferential 
part of the \pkg{PLMIX} package consists of the following three functions, fitting a Bayesian $G$-component PL mixture according to the estimation procedures reviewed in Sections \ref{sss:MAPhete} and \ref{sss:GShete}
\begin{description}[leftmargin=!,labelwidth=\widthof{\code{mapPLMIX\char`_multistart}}]
\item[\code{mapPLMIX}] 
%%% which 
maximizes the posterior 
%%%  distribution 
distribution
via EM algorithm and returns the MAP point estimate of the PL mixture parameters;
\item[\code{mapPLMIX\char`_multistart}] 
%%%  which 
does the same with multiple 
%%% stating 
starting 
values, in order to address the issue of possible local maxima in the posterior distribution;
\item[\code{gibbsPLMIX}] implements the MCMC posterior simulation via GS, 
aimed at quantifying estimation uncertainty from a fully Bayesian perspective.
\end{description}
%
%The main input arguments are listed in the following
%\begin{description}[leftmargin=!,labelwidth=\widthof{\code{mapPLMIX\char`_multistart}}]
%\item[\code{mapPLMIX}] 
%%% which 
%maximizes the posterior distribution via EM algorithm and returns the MAP point estimate of the PL mixture parameters;
%\item[\code{mapPLMIX\char`_multistart}] which does the same with multiple stating values, in order to address the issue of possible local maxima in the posterior distribution.
%\item[\code{gibbsPLMIX}] which implements the MCMC posterior simulation via Gibbs sampling, aimed at quantifying parameter uncertainty from a fully Bayesian perspective.
%\end{description}
%
The above functions can be conveniently applied in a sequential way:
first the MAP procedure can be launched with multiple starting values by using
with \code{mapPLMIX_multistart} and, then, the resulting MAP estimate
can be employed to initialize the MCMC chain in the \code{gibbsPLMIX}
command. 

Since the PL
% reviewed in Section~\ref{ss:pl} 
is parametrized by the item-specific quantities $\underline{p}$
%=(p_1,\dots,p_K)$ 
governing the sequential
% stochastic 
drawings of the items in order of preference,
% during the ranking process. 
%inference on this model requires a set of orderings as input data.
%This implies that 
%a sample 
the ordering format $\underline\pi^{-1}$
% expressed in
% the
%ordering format 
is the natural choice for the input dataset 
of the inferential process.
%to make inference on
%the PL mixture scenario. 
%%% 
For this reason, all the functions concerning model estimation
% ranking analysis 
share the \code{pi_inv} argument, indicating the numeric $N\times K$ matrix of observed partial top orderings.
%For this reason, all 
Here is an example illustrating how to obtain the posterior
mode for a Bayesian 3-component PL mixture fitted to the \code{d_apa}
dataset
%~\citep{Diaconis-Rep} 
under the 
%uninformative 
noninformative 
prior scenario
\begin{CodeChunk}
\begin{CodeInput}
> data(d_apa)
> set.seed(57524)
> MAP_3 <- mapPLMIX_multistart(pi_inv=d_apa, K=5, G=3,
+   n_start=30, n_iter=400*3, centered_start=TRUE, parallel=TRUE)
\end{CodeInput}
\end{CodeChunk}
We run the EM algorithm with \code{n_start=30} starting values
which, if not supplied by the user in the \code{init} argument, are
randomly generated from a uniform distribution (default). The optional \code{centered_start} input is a logical value to constraint the random starting values to be centered around the
observed relative frequency that each item has been ranked
first. Additionally, the \code{hyper} argument contains the
hyperparameters values ($c_{gi}$, $d_g$ and $\alpha_g$) of the
conjugate prior setting arranged in a list of objects named \code{shape0}, \code{rate0} and \code{alpha0}. By default, flat priors are assumed, implying that the MAP estimate
coincides with the MLE solution. From a computational point of view,
note the logical argument 
\code{parallel} 
%%%  that 
%%% gives the possibility to parallelize the different launches 
that allows to parallelize
% parallelized 
the initializations and,
%%% and to 
hence, 
to 
significantly 
%%% reduce 
reduce
the execution time. 

The \code{mapPLMIX_multistart} automatically selects the best solution in terms of maximum value of the posterior distribution and returns a list containing the main information on the implemented MAP procedure. The MAP estimates of the component-specific support parameters and the mixture weights can be extracted by accessing to the corresponding list elements
as follows
\begin{CodeChunk}
\begin{CodeInput}
> MAP_3$mod$P_map
\end{CodeInput}
\begin{CodeOutput}
           [,1]       [,2]       [,3]       [,4]       [,5]
[1,] 0.06247449 0.03295813 0.01664217 0.51188738 0.37603783
[2,] 0.27331708 0.04903217 0.61671929 0.02382562 0.03710584
[3,] 0.18807113 0.22080423 0.14093403 0.22727853 0.22291209
\end{CodeOutput}
\begin{CodeInput}
> MAP_3$mod$W_map
\end{CodeInput}
\begin{CodeOutput}
[1] 0.1035369 0.2732693 0.6231937
\end{CodeOutput}
\end{CodeChunk}
%
%as well as 
%the estimated posterior component membership probabilities and 
The model-based clustering of the sample units into the $G=3$ mixture components based on the MAP allocation
% posterior membership probability that
is recorded in the list element named \code{class_map}. For the \code{d_apa} example, 
%we show in aggregated format
the class distribution turns out to be
\begin{CodeChunk}
\begin{CodeInput}
> table(MAP_3$mod$class_map)
\end{CodeInput}
\begin{CodeOutput}
    1     2     3 
  621  4106 10722 
\end{CodeOutput}
\end{CodeChunk}
%
%As mentioned in Section, the  provides a function to fit 
Notice that a PL mixture can be fitted in \proglang{R} with the function \code{Estimation.RUM.MultiType.MLE} of the \pkg{StatRank} package, by specifying the exponential distribution for the latent random utility. Unfortunately, the long computational time makes the implementation of the PL mixtures unfeasible
%even on a relatively small sample of partial observations.
for a large dataset such as the \code{d_apa}. Indeed, the comparison of the timings elapsed for fitting the PL reported in \cite{PlackettLuce} shows that the \pkg{PLMIX} remarkably outperfoms all the other packages dealing with the PL in terms of computational efficiency.

Subsequently, we can perform an approximation of the posterior distribution by means of the GS simulation
%, taking advantage of the conjugate structure of the model 
%defined in Section \ref{ss:hetero} that is 
implemented in the \code{gibbsPLMIX} command. An example to run the GS initialized with the MAP estimates just obtained from the EM algorithm is 
\begin{CodeChunk}
\begin{CodeInput}
> set.seed(57524)
> GIBBS_3 <- gibbsPLMIX(pi_inv=d_apa, K=5, G=3, init=list(p=MAP_3$mod$P_map,
+   z=binary_group_ind(MAP_3$mod$class_map,G=3)), n_iter=22000,
+   n_burn=2000)
  \end{CodeInput}
\end{CodeChunk}
In the \code{init} argument, the user can provide the list of initial values for the support parameters \code{p} and the binary component membership indicators \code{z}. For the latter, \pkg{PLMIX} offers the  utility \code{binary_group_ind} converting the vector of group labels into the binary matrix $\underline{z}$. If \code{init} values are not supplied, random initialization from the uniform distribution is performed (default). Additionally, \code{n_iter} and \code{n_burn} correspond to the total number of GS drawings and the length of the burn-in phase, implying that the final posterior MCMC sample has size $L=\text{\code{n\char`_iter}}-\text{\code{n\char`_burn}}$. The output is a list of named objects including
% the traces of the support parameters and the weights
the parameter drawings
\begin{CodeChunk}
\begin{CodeInput}
> round(head(GIBBS_3$P), 3)
\end{CodeInput}
\begin{CodeOutput}
      p1,1  p2,1  p3,1  p1,2  p2,2  p3,2  p1,3  p2,3  p3,3
[1,] 0.110 0.656 0.390 0.250 0.119 0.022 0.196 0.030 0.307
[2,] 0.099 0.655 0.339 0.244 0.083 0.022 0.204 0.033 0.427
[3,] 0.097 0.543 0.358 0.260 0.100 0.012 0.192 0.036 0.413
[4,] 0.104 0.647 0.346 0.258 0.083 0.022 0.177 0.032 0.385
[5,] 0.107 0.633 0.336 0.249 0.061 0.032 0.199 0.031 0.337
[6,] 0.114 0.580 0.426 0.237 0.098 0.032 0.201 0.044 0.306
      p1,4  p2,4  p3,4  p1,5  p2,5  p3,5  p1,6  p2,6  p3,6
[1,] 0.199 0.094 0.064 0.078 0.008 0.004 0.168 0.093 0.212
[2,] 0.206 0.107 0.068 0.075 0.008 0.005 0.171 0.114 0.138
[3,] 0.203 0.127 0.060 0.076 0.008 0.004 0.172 0.185 0.153
[4,] 0.207 0.106 0.081 0.077 0.011 0.005 0.176 0.121 0.160
[5,] 0.202 0.102 0.094 0.077 0.012 0.006 0.166 0.161 0.195
[6,] 0.193 0.115 0.057 0.078 0.022 0.005 0.177 0.141 0.173
\end{CodeOutput}
\begin{CodeInput}
> round(head(GIBBS_3$W), 3)
\end{CodeInput}
\begin{CodeOutput}
        w1    w2    w3
[1,] 0.858 0.070 0.072
[2,] 0.888 0.066 0.046
[3,] 0.913 0.045 0.042
[4,] 0.885 0.073 0.042
[5,] 0.868 0.080 0.051
[6,] 0.893 0.051 0.056
\end{CodeOutput}
\end{CodeChunk}
and the posterior likelihood and deviance values at each iteration
\begin{CodeChunk}
\begin{CodeInput}
> head(GIBBS_3$log_lik)
\end{CodeInput}
\begin{CodeOutput}
[1] -2715.182 -2714.959 -2720.991 -2716.492 -2716.751 -2715.400
\end{CodeOutput}
\begin{CodeInput}
> head(GIBBS_3$deviance)
\end{CodeInput}
\begin{CodeOutput}
[1] 5430.365 5429.919 5441.981 5432.983 5433.502 5430.799
\end{CodeOutput}
\end{CodeChunk}
%

%model due to the special form of the associated likelihood. Although lead to a more analytically tractable objective functions, their implementation remains challenging due to demanding computational times. Our Some parts of the code, regarding especially the evaluation of the PL likelihood, have been written in \proglang{C++} to assist the demanding computational times. We exploited the routines in the \pkg{inline} and \pkg{Rcpp} to integrate the \proglang{R} and \proglang{C++}. 

\subsection{Model comparison: APA data}
\label{ss:comparison}
The \code{selectPLMIX} function assists the user in the choice of the number of mixture components 
%returning the value
%\pkg{PLMIX} package computes 
%by means the 
via computation
of the
% Bayesian selection tools 
criteria described in Section~\ref{ss:mc}.
%several Bayesian criteria to address the issue of selecting a suitable number of mixture components. 
Let us suppose that Bayesian PL mixtures have been fitted to the \code{d_apa} dataset with $G$ varying from 1 to 3 with the code just described in Section \ref{ss:estimate}. The comparison of the three estimated mixtures can be performed with the following instruction
\begin{CodeChunk}
\begin{CodeInput}
> SELECT <- selectPLMIX(pi_inv=d_apa, seq_G=1:3, parallel=TRUE,
+   MAPestP=list(MAP_1$mod$P_map, MAP_2$mod$P_map, MAP_3$mod$P_map),
+   MAPestW=list(MAP_1$mod$W_map, MAP_2$mod$W_map, MAP_3$mod$W_map),
+   deviance=list(GIBBS_1$deviance, GIBBS_2$deviance, GIBBS_3$deviance))
\end{CodeInput}
\end{CodeChunk}
Besides the number of components of the competing mixtures specified in the vector \code{seq_G}, the command requires the lists of the point estimates and the posterior \code{deviance} values. More specifically, the function privileges the use of the MAP estimates \code{MAPestP} and \code{MAPestW} but, by setting them to NULL values, the user can alternatively compute the selection measures by relying on the a different posterior summary (\code{"mean"} or \code{"median"}) specified in the \code{post_summary} argument. In the latter case, the command needs also the MCMC samples to compute the desired posterior summary, that have to be supplied in the \code{MCMCsampleP} and \code{MCMCsampleW} arguments. The drawback when working with point estimates other than the MAP is that the presence of LS has to be previously removed from the traces to obtain meaningful results. Notice also the \code{parallel} option to parallelize the computation over the alternative number of groups specified in the \code{seq_G} argument.
%The output is a list with detailed information not only on the comparison tools, but also on the separate terms composing them, namely the fitting measures (\code{} and \code{}) and the effective number of parameters (\code{} and \code{}). 
The final values of the criteria can be extracted by typing
\begin{CodeChunk}
\begin{CodeInput}
> SELECT$selection_criteria
\end{CodeInput}
\begin{CodeOutput}
        DIC1     DIC2    BPIC1    BPIC2    BICM1    BICM2
G=1 103204.4 103204.3 103208.3 103208.0 103233.0 103232.9
G=2 100771.9 100772.7 100779.4 100780.9 100835.6 100836.3
G=3 100591.1 100593.0 100601.3 100605.1 100685.7 100687.6
\end{CodeOutput}
\end{CodeChunk}
In this example, the decreasing trend of all the measures clearly 
% suggest 
suggests that more complex mixtures with additional components should be explored. Finally, in the case of an uninformative analysis, a comparison with the frequentist solution is allowed. In this regard, the BIC value is returned by the \code{mapPLMIX_multistart} when flat priors are adopted. For the three mixtures, one has
\begin{CodeChunk}
\begin{CodeInput}
> rbind(MAP_1$mod$bic, MAP_2$mod$bic, MAP_3$mod$bic)
\end{CodeInput}
\begin{CodeOutput}
         [,1]
[1,] 5475.685
[2,] 5484.724
[3,] 5504.845
\end{CodeOutput}
\end{CodeChunk}
Alternatively, the computation of the BIC can be accomplished with the \code{bicPLMIX} utility which, similarly to the \code{loglikPLMIX} function, accommodates for the more general EPL mixture setting.
% with the additional reference order parameters.
%The use of WAIC is strongly recommended in practice by \cite{Gelman:Hwang:Vehtari}, because of its connections with the Bayesian cross-validation
%%, formally proven in 
%\citep{Watanabe2010} and its validity for singular models (it does not relies on point estimates). On the other hand, its computation is particularly time-consuming as it requires a pointwise calculation of the log-likelihood. 
%For this reason, we have written a dedicate function in \proglang{C++} to significantly reduce the runtime, with a great improvement over other \proglang{R} commands assisting the implementation of WAIC, such as \code{WAIC} in the \pkg{blmeco} package and \code{waic} in the \pkg{loo} (FAI UN CONFRONTO, SE POSSIBILE).

\subsection{Model assessment: APA data}
\label{ss:assess}
The posterior predictive check, unconditionally and conditionally on
the 
%number of ranked items,
length of the partial sequences, can be performed,
 %has been implemented 
respectively, with the
\code{ppcheckPLMIX} and \code{ppcheckPLMIX_cond} functions. As
described in Section~\ref{ss:ma}, the model assessment tools require
the simulation of a replicated dataset from the posterior predictive
distribution for each
% parameter value of the 
GS drawing.
%simulation. 
This means
that the execution time depends on both the sample sizes $N$ and $L$
and, hence, the computation of goodness-of-fit diagnostics is
particularly time-consuming. Thanks to the combination of the
\proglang{R}
% environment with the 
and \proglang{C++} languages, the assessment
of ranking models becomes feasible with the \pkg{PLMIX} package,
%%%
even for moderately large datasets. 
The code to perform the posterior predictive check based on 
%the chi-squared discrepancies \eqref{e:chi} 
$ X^2_{(1)}$ and $ X^2_{(2)}$ and to extract the corresponding $p$-values is
\begin{CodeChunk}
\begin{CodeInput}
> set.seed(57524)
> CHECK <- ppcheckPLMIX(pi_inv=d_apa, seq_G=1:3, parallel=TRUE,
+   MCMCsampleP=list(GIBBS_1$P, GIBBS_2$P, GIBBS_3$P),
+   MCMCsampleW=list(GIBBS_1$W, GIBBS_2$W, GIBBS_3$W))
> CHECK$post_pred_pvalue
\end{CodeInput}
\begin{CodeOutput}
    post_pred_pvalue_top1 post_pred_pvalue_paired
G_1                     0                  0.0000
G_2                     0                  0.6330
G_3                     0                  0.4805
\end{CodeOutput}
\end{CodeChunk}
The syntax is similar to that shown for the \code{selectPLMIX} command, with the difference that the lists \code{MCMCsampleP} and \code{MCMCsampleW} collecting the MCMC samples are necessary inputs for the posterior predictive simulation.
Similarly, the script for the conditional posterior predictive check based on 
%the chi-squared discrepancies \eqref{e:chitilde} 
$\tilde X^2_{(1)}$ and $\tilde X^2_{(2)}$ is 
\begin{CodeChunk}
\begin{CodeInput}
> set.seed(57524)
> CHECKCOND <- ppcheckPLMIX_cond(pi_inv=d_apa, seq_G=1:3, parallel=TRUE,
+   MCMCsampleP=list(GIBBS_1$P, GIBBS_2$P, GIBBS_3$P),
+   MCMCsampleW=list(GIBBS_1$W, GIBBS_2$W, GIBBS_3$W))
> CHECKCOND$post_pred_pvalue
\end{CodeInput}
\begin{CodeOutput}
    post_pred_pvalue_top1_cond post_pred_pvalue_paired_cond
G_1                          0                            0
G_2                          0                            0
G_3                          0                            0
\end{CodeOutput}
\end{CodeChunk}
Remind that under correct model specification, $p_B$ values are
expected to be 
%%% 
centered
around 0.5, whereas values 
%of $p_B$ 
smaller than 0.05  are typically considered as indication of model
lack-of-fit. 
%From the results of the unconditional analysis, it
%follows that none of the three mixtures 
%provides an adequate description of the observed frequencies that each item has been ranked first. Regarding the PC frequencies, instead, only the homogeneous PL is not able to recover this bivariate feature of the entire observed sample.
%Interestingly, 
In this example, the posterior predictive check conditionally on the
number of ranked items reveals the inadequacy of all estimated
mixtures for both the summary statistics. This should be interpreted
as an indication that a better account of the missingness mechanism 
is
% need 
needed 
and, hence, a separate PL mixture analysis on each subsample $\upi^{-1}_m$
%of partial orderings
% with the same length 
would be preferable. See \cite{Mollica:Tardella2017} for a more in-depth analysis of the \code{d_apa} dataset.

\subsection{Label switching adjustment: simulated data}
\label{ss:ls}
The \code{label_switchPLMIX} command can be employed to remove the possible presence of LS in the posterior MCMC samples. This step is necessary to derive meaningful point estimates 
%of the PL mixture parameters 
other than the MAP and the related uncertainty measures. The
% aforementioned 
function relies on the application of the Pivotal Reordering Algorithm (PRA) proposed by~\cite{Marin:Mengersen:Robert} by means of a call to the \code{pra} routine of the R package \code{label.switching}~\citep{Papastamoulis}. 
%The PRA acts in an intuitive way: it switches the component labels of each GS drawing
%according to a permutation such that a certain distance from a target posterior mode is minimized. The target mode plays the role of \textit{pivot} and can be easily identified with the MAP solution.

To illustrate the LS adjustment, we first generated a sample of $N=300$ orderings of $K=4$ items from a 2-component PL mixture
% and the following parameter configuration
%
\begin{CodeChunk}
\begin{CodeInput}
> p_par <- rbind(c(.7,.2,.08,.02), c(.55,.3,.03,.12))
> w_par <- c(0.7, 0.3)
> set.seed(70476)
> sim_orderings <- rPLMIX(n=300, K=4, G=2, p=p_par,
+   weights=w_par, format="ordering)$sim_data
\end{CodeInput}
\end{CodeChunk}
%
%The synthetic data in aggregated format are
%%
%\begin{CodeChunk}
%\begin{CodeInput}
%> unit_to_freq(sim_orderings)
%\end{CodeInput}
%\begin{CodeOutput}
%      [,1] [,2] [,3] [,4] [,5]
% [1,]    1    2    3    4   87
% [2,]    1    2    4    3   52
% [3,]    1    3    2    4   40
% [4,]    1    3    4    2    1
% [5,]    1    4    2    3   13
% [6,]    1    4    3    2    8
% [7,]    2    1    3    4   36
% [8,]    2    1    4    3   20
% [9,]    2    3    1    4    9
%[10,]    2    4    1    3    5
%[11,]    2    4    3    1    1
%[12,]    3    1    2    4    8
%[13,]    3    2    1    4    3
%[14,]    3    2    4    1    1
%[15,]    4    1    2    3    9
%[16,]    4    1    3    2    4
%[17,]    4    2    1    3    3
%\end{CodeOutput}
%\end{CodeChunk}
%%
%%
%\begin{CodeChunk}
%\begin{CodeInput}
%> modal_orderings <- t(apply(p_par,1, order, decreasing=TRUE))
%> modal_orderings
%\end{CodeInput}
%\begin{CodeOutput}
%     [,1] [,2] [,3] [,4]
%[1,]    1    2    3    4
%[2,]    1    2    4    3
%\end{CodeOutput}
%\end{CodeChunk}
%%
%Note that, 
With this
% support 
parameter setting, the component-specific
modal orderings turn out to be adjacent
% on the ranking 
% pace
%space 
in terms of the Kendall distance, since only their last two positions are switched. Of course, the closeness of the PL components facilitates the occurrence of LS.
Then, we fitted the 2-PL mixture with uninformative priors
% densities 
by means of
%%first recorded the MAP estimate 
%$\text{\code{n\char`_start}}=30$ parallelized launches of 
the EM algorithm
%, resulting
%% and the best fitting model returned 
%in the following MAP estimates
and finally we used the resulting MAP solutions to initialize the GS
\begin{CodeChunk}
\begin{CodeInput}
> set.seed(70476)
> MAP <- mapPLMIX_multistart(pi_inv=sim_orderings, K=4, G=2,
+   n_start=30, n_iter=1000, parallel=TRUE)
MAP$mod$P_map
\end{CodeInput}
\begin{CodeOutput}
          [,1]        [,2]        [,3]       [,4]
[1,] 0.6535795 0.252212857 0.061412559 0.03279511
[2,] 0.5023897 0.001954253 0.002959641 0.49269641
\end{CodeOutput}
\begin{CodeInput}
MAP$mod$W_map
\end{CodeInput}
\begin{CodeOutput}
[1] 0.96190009 0.03809991
\end{CodeOutput}
\begin{CodeInput}
> set.seed(70476)
> GIBBS <- gibbsPLMIX(pi_inv=sim_orderings, K=4, G=2, ,
+   init=list(p=MAP$mod$P_map, z=binary_group_ind(MAP$mod$class_map, G=2)))
\end{CodeInput}
\end{CodeChunk}
%
%obtained from
%Finally, we used the MAP solutions
%that, subsequently, were employed 
%to initialize the GS 
%%
%\begin{CodeChunk}
%\begin{CodeInput}
%> set.seed(70476)
%> GIBBS <- gibbsPLMIX(pi_inv=sim_orderings, K=4, G=2, ,
%+   init=list(p=MAP$mod$P_map, z=binary_group_ind(MAP$mod$class_map, G=2)))
%\end{CodeInput}
%\end{CodeChunk}
%%
%%
\begin{figure}[t]
\centering
\includegraphics[scale=.6]{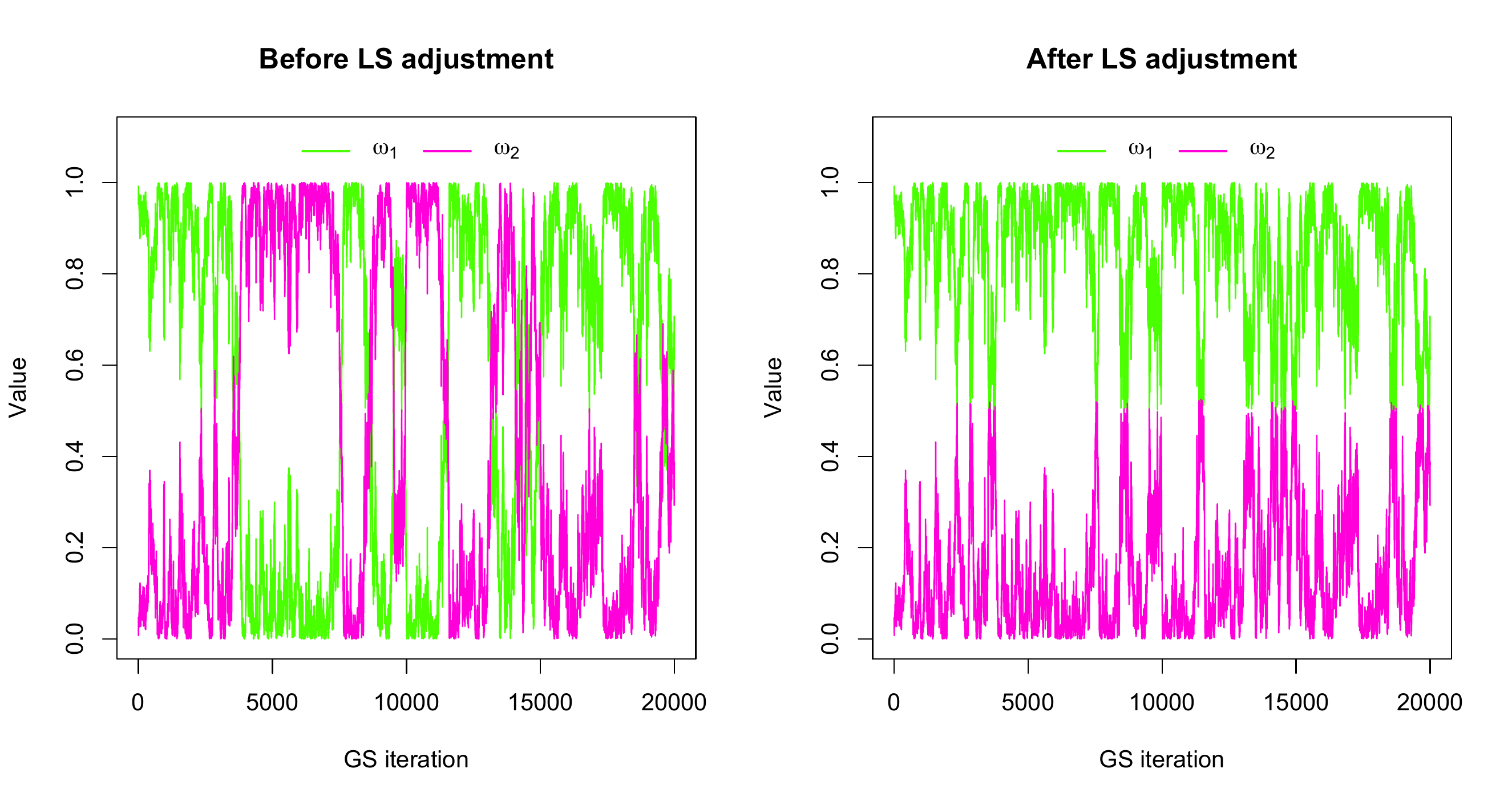}
\caption{Traceplots of mixture weights before and after the application of the PRA.}
\label{fig:LS}
\end{figure}
%
%The first $\text{\code{n\char`_burn}}=2000$) drawings were discarded as burn-in period from a total of $\text{\code{n\char`_iter}}=22000$ iterations.
%and show the overlapped 
%The traceplots of the two mixture weight samples 
The two samples of the mixture weights
are shown
% are shown 
in Figure~\ref{fig:LS} (left).
% and reveal the clear 
The occurrence of LS is
testified by the multiple swaps of the traceplots, 
%evident and manifests 
%
%itself
%with step-like configurations of the traceplots,
indicating several transitions of the sampler from an artificial mode to another. %caused by the identifiability issue.
Indeed, a remarkable percentage of the chain
% posterior sample 
is affected by LS, leading to similar (and invalid) %marginal 
posterior means.
Here are
% see for example
% the point estimates 
those of the support parameters
\begin{CodeChunk}
\begin{CodeInput}
> matrix(colMeans(GIBBS$P), ncol=4)
          [,1]      [,2]       [,3]       [,4]
\end{CodeInput}
\begin{CodeOutput}
[1,] 0.5761867 0.2621617 0.07650548 0.08514619
[2,] 0.5415158 0.2795709 0.08259536 0.09631793
\end{CodeOutput}
\end{CodeChunk}
%

%Now we 
%can try to remove the LS phenomenon by 
The post-processing of the raw MCMC output with the PRA
%, aimed at removing the LS phenomenon,
%. To this aim, some default input values of the \code{ppcheckPLMIX} and \code{ppcheckPLMIX_cond} commands have to be modified, as indicated in 
can be implemented as follows 
%with the following instruction
%
\begin{CodeChunk}
\begin{CodeInput}
> LS <- label_switchPLMIX(pi_inv=sim_orderings, seq_G=2,
+   MCMCsampleP=list(GIBBS$P), MCMCsampleW=list(GIBBS$W),
+   MAPestP=list(MAP$mod$P_map), MAPestW=list(MAP$mod$W_map))
\end{CodeInput}
%\begin{CodeOutput}
%[1] "LABEL SWITCHING ADJUSTMENT WITH PIVOTAL REORDERING ALGORITHM"
%\end{CodeOutput}
\end{CodeChunk}
whose inputs values are defined as those of the functions for the model selection and the posterior predictive check.
%\code{selectPLMIX}, \code{ppcheckPLMIX} and \code{ppcheckPLMIX_cond} functions.
%The argument \code{ls_adj=TRUE} has to specified to request the adjustment of the GS simulation 
%%from the posterior distribution 
%with the PRA, whereas \code{adj_post_sample=TRUE} is needed to record the relabeled posterior samples in the output list with names \code{final_sampleP} and \code{final_sampleW}.
%% for the sets of PL mixture parameters. 
%With the logical \code{top1} and \code{paired} arguments equal to \code{FALSE}, the function skips the computation of the posterior predictive diagnostics.
The traceplots in Figure~\ref{fig:LS} (right) reveal a very good performance of the PRA in removing the artificial multimodality. The adjusted
% allowing for the computation of 
posterior summaries are
\begin{CodeChunk}
\begin{CodeInput}
> apply(LS$final_sampleP$G_2, 2, rowMeans)
\end{CodeInput}
\begin{CodeOutput}
          [,1]      [,2]       [,3]       [,4]
[1,] 0.6742229 0.2278699 0.06266152 0.03524561
[2,] 0.4434795 0.3138627 0.09643932 0.14621852
\end{CodeOutput}
\begin{CodeInput}
> colMeans(LS$final_sampleW$G_2)
\end{CodeInput}
\begin{CodeOutput}
[1] 0.8511395 0.1488605
\end{CodeOutput}
\end{CodeChunk}
%Finally, 
We can 
% note 
%still 
detect
a certain discrepancy between the adjusted GS estimates and the true parameter values, although the actual order of the support parameters within each group is fully recovered.
%Table \ref{t:ARISimul} contains the ARI values comparing the model-based classifications with the true one.
As expected, when the two mixture components considerably overlap, it is more difficult to reconstruct the actual group membership of the sample units,
% and this can have
with consequent negative effects on the final estimates.
%The ARI is remarkably poor in the third Scenario due to the
On the other hand, the performance of the GS turns out to be better than the MAP estimate
% corresponding to the 
(MLE solution), since the latter completely fails to infer the minor mixture component. 
\subsection{A comparison with the prefmod package: CARCONF data}
\label{ss:apprealCARCONF}
%
%In this section we show the use of the main \pkg{PLMIX} functions with an application of the Bayesian PL mixtures to the real dataset \code{d_carconf}. 
To further highlight the possible advantages of the \pkg{PLMIX} package, a comparison with some methods implemented in the \proglang{R} package \pkg{prefmod}~\citep{Hatz:Ditt:2012} is provided. \pkg{prefmod} represents a
%one of the most recent and 
flexible package
% allowing 
for the analysis of preference data expressed in the form of 
%rankings through extensions of the BTs for PC data.
PCs. The same framework is applicable also for ranking data. A ranking of $K$ items, in fact, can be 
%equivalently expressed in terms of the corresponding set of 
decomposed into the equivalent \textit{pattern} of $K(K-1)/2$ PCs, where the alternatives are compared two at a time and the preferred one is specified. For this reason, ranking models based on PCs are also referred to as \textit{pattern models}.

To explore the unobserved sample heterogeneity of the CARCONF data with the \pkg{prefmod} package, we considered the nonparametric maximum likelihood approach (NPML) described in~\citep{Hatz:Ditt:2012} and estimated pattern models with discrete random effects. In this way, the resulting NPML clustering of the sample units into latent classes can be more straightforwardly compared with the classification via finite PL mixtures.
Since the NPML method in \pkg{prefmod} accepts only full observations as input data, we first performed a completion of the partial ordering dataset \code{d_carconf} with the function \code{make_complete}, by using the frequencies $\underline{r}(\upi^{-1})$ for the random imputation
% of the missing items as follows

%
\begin{CodeChunk}
\begin{CodeInput}
> N <- nrow(d_carconf)
> K <- ncol(d_carconf)
> summaries <- rank_summaries(data=d_carconf_compl, format="ordering", 
+   mean_rank=FALSE, pc=TRUE)
> top_freq <- summaries$marginals["Rank_1",]
> set.seed(57524)
> d_carconf_compl <- make_complete(data=d_carconf, format="ordering", 
+   probitems=top_freq)$completedata
\end{CodeInput}
\end{CodeChunk} 
and we then converted the dataset into a \code{data.frame} of rankings with labeled columns denoting the $K=6$ car modules
\begin{CodeChunk}
\begin{CodeInput}
> d_carconf_compl_r <- data.frame(rank_ord_switch(d_carconf_compl,
+   format="ordering")) 
> names(d_carconf_compl_r) <- c("price", "exterior", "brand",
+   "tech.equip", "country", "interior")
\end{CodeInput}
\end{CodeChunk} 
After constructing the design matrix needed for the \pkg{prefmod} commands
%associated to the dataset, obtained with the following instructions,
%
\begin{CodeChunk}
\begin{CodeInput}
> library(prefmod)
> dsg <- patt.design(obj=d_carconf_compl_r, nitems=K,
+   objnames=names(d_carconf_compl_r), resptype="ranking")
\end{CodeInput}
\end{CodeChunk} 
four random effects pattern models were estimated
% with the NPML method 
with the function \code{pattnpml.fit}, by varying the number of latent classes from $G=1$ to $G=4$
\begin{CodeChunk}
\begin{CodeInput}
> npml1 <- pattnpml.fit(formula= y ~ price + exterior + brand +
+   tech.equip + country + interior, k=1, design=dsg, seed=57524)
> npml2 <- pattnpml.fit(formula= y ~ 1, random= ~price + exterior + brand +
+   tech.equip + country + interior, k=2, design=dsg, seed=57524)
> npml3 <- update(npml2, k=3)
> npml4 <- update(npml2, k=4)
\end{CodeInput}
\end{CodeChunk} 
The corresponding BIC values are listed below
\begin{CodeChunk}
\begin{CodeInput}
> BIC(npml1, npml2, npml3, npml4)
\end{CodeInput}
\begin{CodeOutput}
      df      BIC
npml1  6 1385.398
npml2 12 1398.626
npml3 18 1431.977
npml4 24 1468.038
\end{CodeOutput}
\end{CodeChunk} 
suggesting the homogeneous ($G=1$) pattern model as the optimal one (minimum BIC value).
For comparison purposes, we re-fitted the selected 1-class pattern model within the MLE framework and computed the corresponding BIC
% with the following code
%
\begin{CodeChunk}
\begin{CodeInput}
> patt.mod <- pattR.fit(obj=d_carconf_compl_r, nitems=K,
+   obj.names=names(d_carconf_compl_r))
> -2*patt.mod$ll + (K-1)*log(N)
\end{CodeInput}
\begin{CodeOutput}
[1] 5509.968
\end{CodeOutput}
\end{CodeChunk} 
%
%leading to support parameter estimates equal to
%%
%\begin{CodeChunk}
%\begin{CodeInput}
%> worthPATT <- patt.worth(patt.mod)
%> worthPATT
%\end{CodeInput}
%\begin{CodeOutput}
%            estimate
%price      0.1556699
%exterior   0.1948305
%brand      0.1823920
%tech.equip 0.1796178
%country    0.1138552
%interior   0.1736346
%attr(,"class")
%[1] "wmat"   "matrix"
%\end{CodeOutput}
%\end{CodeChunk} 
%%
%and computed the corresponding BIC, equal to
%%
%\begin{CodeChunk}
%\begin{CodeInput}
%> -2*patt.mod$ll + (K-1)*log(N)
%\end{CodeInput}
%\begin{CodeOutput}
%[1] 5509.968
%\end{CodeOutput}
%\end{CodeChunk} 
%%
By adopting the MAP procedure with flat priors to fit $G$-component PL mixtures with $G=1,\dots,4$ and to recover the MLE solutions, we obtained the following BIC results
\begin{CodeChunk}
\begin{CodeInput}
> for(i in 1:4){
+   set.seed(57524)
+   assign(paste0("MAP_",i), mapPLMIX_multistart(pi_inv=d_carconf_compl, K=K,
+     G=i, n_start=30, n_iter=400*i, parallel=TRUE))
+ }
> rbind(MAP_1$mod$bic, MAP_2$mod$bic, MAP_3$mod$bic, MAP_4$mod$bic)
\end{CodeInput}
\begin{CodeOutput}
         [,1]
[1,] 5475.685
[2,] 5484.724
[3,] 5504.845
[4,] 5530.541
\end{CodeOutput}
\end{CodeChunk} 
Interestingly, the minimum BIC value is still achieved in correspondence of the homogeneous model, but it turns out to be significantly smaller than that associated to the pattern model, meaning that the PL assumption considerably improves the fitting of the CARCONF data. To stress the importance of goodness-of-fit diagnostics,
% for ranking data analysis, 
we also checked the adequacy of the two frequentist models to recover the sample statistics described in Section~\ref{ss:ma},
%whose observed values are 
given by 
\begin{CodeChunk}
\begin{CodeInput}
> top_freq <- rank_summaries(data=d_carconf_compl, format="ordering", 
+   mean_rank=FALSE, pairedcomparisons=TRUE)$marginals["Rank_1",]
> pc_freq <- summaries$pairedcomparisons
> pc_freq <- pc_freq[lower.tri(pc_freq)]
\end{CodeInput}
%\begin{CodeOutput}
% [1] 261 256 253 157 245 197 196  93 189 205 111 209 112 201 318
%\end{CodeOutput}
\end{CodeChunk} 
By adopting the traditional chi-squared test, for the 1-class pattern model we obtained
\begin{CodeChunk}
\begin{CodeInput}
> worthPATT <- patt.worth(patt.mod)
> chisq.test(x=top_freq, p=c(worthPATT), correct=FALSE, rescale.p=TRUE)
\end{CodeInput}
\begin{CodeOutput}
[1] 0.000244741
\end{CodeOutput}
\begin{CodeInput}
> df2 <- K*(K-1)/2-1
> n.tot.matches=rep(N,df2+1)
> exp.freq.pcPATT=Freq_th(p=worthPATT,n.matches=n.tot.matches)[,2]
> obs.chisq.pcPATT <- sum((pc_freq-exp.freq.pcPATT)^2/exp.freq.pcPATT)
> pchisq(q=obs.chisq.pcPATT, df=df2, lower.tail=FALSE)
\end{CodeInput}
\begin{CodeOutput}
[1] 1.39225e-13
\end{CodeOutput}
\end{CodeChunk} 
where the function \code{patt.worth} returns the estimated support parameters of the pattern model, needed for the computation of the expected frequencies. As evident, both $p$-values are well below the critical threshold 0.05, indicating a remarkably poor fitting.
However, some deficiencies to recover the marginal most-liked item distribution can be highlighted also for the 1-component PL mixture, whereas they do not seem to emerge for the PCs
% This is proved by the following results
%%
%\begin{CodeChunk}
%\begin{CodeInput}
%> MAP_1$mod$P_map
%\end{CodeInput}
%\begin{CodeOutput}
%          [,1]      [,2]      [,3]      [,4]       [,5]      [,6]
%[1,] 0.1252602 0.2327074 0.1949075 0.1957678 0.07023363 0.1811234
%\end{CodeOutput}
%\begin{CodeInput}
%> exp.freqPL <- N*MAP_1$mod$P_map
%> round(exp.freqPL, 0)
%\end{CodeInput}
%\begin{CodeOutput}
%     [,1] [,2] [,3] [,4] [,5] [,6]
%[1,]   54  101   85   85   31   79
%\end{CodeOutput}
%\begin{CodeInput}
%> obs.chisq.topPL <- sum((top_freq-exp.freqPL)^2/exp.freqPL) # 26.27986
%> obs.chisq.topPL
%\end{CodeInput}
%\begin{CodeOutput}
%[1] 26.27986
%\end{CodeOutput}
%\begin{CodeInput}
%> pchisq(q=obs.chisq.topPL, df=df1, lower.tail=FALSE) # 7.874663e-05
%\end{CodeInput}
%\begin{CodeOutput}
%[1] 7.874663e-05
%\end{CodeOutput}
%\begin{CodeInput}
%> exp.freq.pcPL=Freq_th(p=MAP_1$mod$P_map,n.matches=n.tot.matches)[,2]
%> c(exp.freq.pcPL, 0)   
%\end{CodeInput}
%\begin{CodeOutput}
% [1] 282.7845 264.8136 265.2697 156.2792 257.1569 198.2737 198.7490 100.8501
% [9] 190.3886 217.9790 115.2278 209.5271 114.8551 209.0488 313.4533   0.0000
%\end{CodeOutput}
%\begin{CodeInput}
%> obs.chisq.pcPL <- sum((pc_freq-exp.freq.pcPL)^2/exp.freq.pcPL)
%> obs.chisq.pcPL
%\end{CodeInput}
%\begin{CodeOutput}
%[1] 5.160525
%\end{CodeOutput}
%\begin{CodeInput}
%> pchisq(q=obs.chisq.pcPL, df=df2, lower.tail=FALSE)
%\end{CodeInput}
%\begin{CodeOutput}
%[1] 0.9834508
%\end{CodeOutput}
%\end{CodeChunk} 
%%
%%
\begin{CodeChunk}
\begin{CodeInput}
> chisq.test(x=top_freq, p=c(MAP_1$mod$P_map), correct=FALSE, rescale.p=TRUE)
\end{CodeInput}
\begin{CodeOutput}
[1] 0.000244741
\end{CodeOutput}
\begin{CodeInput}
> exp.freq.pcPL=Freq_th(p=MAP_1$mod$P_map,n.matches=n.tot.matches)[,2]
> obs.chisq.pcPL <- sum((pc_freq-exp.freq.pcPL)^2/exp.freq.pcPL)
> pchisq(q=obs.chisq.pcPL, df=df2, lower.tail=FALSE)
\end{CodeInput}
\begin{CodeOutput}
[1] 0.9834508
\end{CodeOutput}
\end{CodeChunk} 
We finally estimated Bayesian PL mixtures up to $G=4$ components
% within a fully Bayesian approach 
by means of the GS algorithm. The MCMC chains were initialized with the MAP solutions and a sample of $\text{\code{n\char`_iter}}=22000$ drawings was obtained for each mixture, including a burn-in phase of $\text{\code{n\char`_burn}}=2000$ iterations
\begin{CodeChunk}
\begin{CodeInput}
> for(i in 1:4){
+   set.seed(57524)
+   assign(paste0("GIBBS_",i), gibbsPLMIX(pi_inv=d_carconf_compl, K=K, G=i, 
+     init=list(p=get(paste0("MAP_",i))$mod$P_map,
+     z=binary_group_ind(get(paste0("MAP_",i))$mod$class_map,G=i)), 
+     n_iter=22000, n_burn=2000))
+ }
\end{CodeInput}
\end{CodeChunk} 
The Bayesian model selection criteria are equal to
\begin{CodeChunk}
\begin{CodeInput}
> selectPLMIX(pi_inv=d_carconf_compl, seq_G=1:4,
+   MAPestP=list(MAP_1$mod$P_map, MAP_2$mod$P_map,
+   MAP_3$mod$P_map, MAP_4$mod$P_map),
+   MAPestW=list(MAP_1$mod$W_map, MAP_2$mod$W_map,
+   MAP_3$mod$W_map, MAP_4$mod$W_map),
+   deviance=list(GIBBS_1$deviance, GIBBS_2$deviance,
+   GIBBS_3$deviance, GIBBS_4$deviance))$selection_criteria
\end{CodeInput}
\begin{CodeOutput}
        DIC1     DIC2    BPIC1    BPIC2    BICM1    BICM2
G_1 5455.352 5455.504 5460.374 5460.678 5476.590 5476.742
G_2 5442.707 5442.993 5455.114 5455.684 5494.715 5495.000
G_3 5443.550 5445.584 5464.543 5468.612 5539.429 5541.463
G_4 5453.448 5446.901 5484.768 5471.674 5547.859 5541.312
\end{CodeOutput}
\end{CodeChunk} 
where, with the exception of BICMs, minimal values are reached by the 2-component PL mixture. The evidence in favour of unobserved sample heterogeneity is reinforced by the posterior predictive $p$-values, given by
\begin{CodeChunk}
\begin{CodeInput}
> set.seed(57524)
> ppcheckPLMIX(pi_inv=d_carconf_compl, seq_G=1:4,
+   MCMCsampleP=list(GIBBS_1$P, GIBBS_2$P, GIBBS_3$P, GIBBS_4$P),
+   MCMCsampleW=list(GIBBS_1$W, GIBBS_2$W, GIBBS_3$W, GIBBS_4$W),
+   parallel=TRUE)$post_pred_pvalue
\end{CodeInput}
\begin{CodeOutput}
    post_pred_pvalue_top1 post_pred_pvalue_paired
G_1               0.00025                 0.33580
G_2               0.09930                 0.51095
G_3               0.10675                 0.50740
G_4               0.10385                 0.49965
\end{CodeOutput}
\end{CodeChunk} 
that, for $G>1$, are all above the reference threshold 0.05. 
%Thus, by following the recommendations in~\cite{Mollica:Tardella2017} in support of DIC and BPIC results, 
The posterior samples for the optimal Bayesian 2-component PL mixture can be finally adjusted to remove the LS 
\begin{CodeChunk}
\begin{CodeInput}
> LS <- label_switchPLMIX(pi_inv=d_carconf_compl, seq_G=2,
+   MCMCsampleP=list(GIBBS_2$P), MCMCsampleW=list(GIBBS_2$W),
+   MAPestP=list(MAP_2$mod$P_map), MAPestW=list(MAP_2$mod$W_map))

\end{CodeInput}
\end{CodeChunk} 
The final posterior means and standard deviations can be easily computed as follows
\begin{CodeChunk}
\begin{CodeInput}
> round(colMeans(LS$final_sampleW$G_2), 3)             				
\end{CodeInput}
\begin{CodeOutput}
[1] 0.769 0.231
\end{CodeOutput}
\begin{CodeInput}
> round(apply(LS$final_sampleP$G_2, 2, rowMeans), 3)             				
\end{CodeInput}
\begin{CodeOutput}
      [,1]  [,2]  [,3]  [,4]  [,5]  [,6]
[1,] 0.092 0.261 0.182 0.195 0.069 0.201
[2,] 0.445 0.113 0.176 0.135 0.041 0.090
\end{CodeOutput}
\begin{CodeInput}
> round(apply(LS$final_sampleW$G_2, 2, sd), 3)             				
\end{CodeInput}
\begin{CodeOutput}
[1] 0.098 0.098
\end{CodeOutput}
\begin{CodeInput}
> round(apply(LS$final_sampleP$G_2, c(1,2), sd), 3)             				
\end{CodeInput}
\begin{CodeOutput}
      [,1]  [,2]  [,3]  [,4]  [,5]  [,6]
[1,] 0.014 0.019 0.016 0.014 0.007 0.020
[2,] 0.124 0.039 0.059 0.036 0.018 0.027
\end{CodeOutput}
\end{CodeChunk} 

\section{Conclusions}
\label{s:concl}
%
%
%Indeed, random data generation is the first acid test for the estimation procedures of novel models, aimed at showing their differential utility over the competing distributional alternatives. 

%the novelties introduced by the \pkg{PLMIX} package can be motivated from several perspectives:
%%
%\begin{itemize}
%\item[-] it contributes to fill the gap concerning the Bayesian estimation of ranking models in \proglang{R} focusing on the Plackett-Luce model and its extensions as generative distribution;
%%\pkg{StatMethRank} package is limited to the Bayesian Multivariate Normal OS model described \cite{Yu} without the mixture configuration and for complete sequences only;
%\item[-] it addresses this issue of  computationally demanding procedures
%%related to the inference on the PL parameters 
%by combining the \proglang{R} environment with \proglang{C++} code and including the parallelization option to simultaneously analyze finite mixtures with a different number of components;
%%\item[-] it involves the parallelization for the estimation of PL mixture with alternative number of components;
%%\item[-] it implements several model comparison criteria and diagnostic tools for goodness-of-fit evaluation;
%\item[-] it covers the fundamental phases of ranking data analysis
%  allowing for a more careful and critical application of ranking
%  models in 
%%%% the 
%real experiments.
%\end{itemize}
%%

When approaching a ranking data analysis, several issues may arise, mainly due to the peculiar structure of ranked sequences as multivariate ordinal data. First, ranking data take values in the finite discrete set of permutations, whose size $K!$ explodes with the total number of items to be ranked. In this perspective, some related difficulties 
%related to the cardinality of the ranking space 
are the possible occurrence of sparse data situations or the need of a manageable exploration of the ranking space.
%, especially when the adopted model includes permutation parameters,
%% in its parametric form, 
%such as the DB and the EPL. 
Secondly, the presence of partial
observations adds further complications. 
%RIVEDERE QUESTA FRASE IN RELAZIONE AI COMMENTI DELLA VIGNETTA 
When the sample is composed
of complete sequences, in fact, the PL could be estimated with methods
related to the log-linear models~\citep{Fienberg} but, in the case of partial
orderings, the likelihood has to be suitably updated and the existing
methods are no longer applicable. All these issues lead to
computationally demanding methods and to the need of developing
specialized softwares to avoid prohibitive execution time. On the
other hand, this has been traditionally 
% prevented from 
an obstacle for a wider 
use of more sophisticated models.  

In order to efficiently address the aforementioned practical issues
and 
% enlarge the application of new 
promote the effective 
% implementation 
exploration of methodological 
advances,
% approaches in practice, 
we developed the 
%novel 
\proglang{R} package \pkg{PLMIX}. This is the first package in \proglang{R} that implements
% promoting 
%the 
ranking data analysis from a Bayesian inferential perspective. It relies on a hybrid code combining \proglang{R} and \proglang{C++} and exploits the advantages of both programming languages, in particular the flexibility of the \proglang{R} environment and the speed of compiled \proglang{C++} code to guarantee computational efficiency. \pkg{PLMIX} is not limited to inferential techniques but represents a comprehensive toolkit, paying special attention to each step of the 
%model-based classification
%% of heterogeneous partial rankings/orderings 
%via a 
Bayesian PL mixture approach to identify clusters of rankers with similar preferences. 
In this regard, the comparative application in Section~\ref{ss:apprealCARCONF} motivates the effectiveness of the Bayesian PL mixture as a profitable parametric alternative 
%and, hence, the usefulness of \pkg{PLMIX} to enlarge the range of options 
for model-based clustering of partial ranking data in \proglang{R}, and the usefulness of the novel goodness-of-fit diagnostics introduced by \cite{Mollica:Tardella2017}.
%Similarly to \cite{Caron:Doucet}, also \cite{Mollica:Tardella2017} detail the MAP estimation and the GS algorithm to perform a fully Bayesian inference on the finite PL mixture. 
%improvements over existing packages. Attention is payed to all the step of the mixture model analysis. 
%Regarding the latter point, the novel package implements the recent model assessment tools described by \cite{Mollica:Tardella2017}, that can be performed unconditionally and conditionally to the length of the partial orderings to better highlight possible deficiencies of the fitted model.
% to reproduce specific features of the observed data. 
%Finally, some examples showed that the execution of \pkg{PLMIX} functions turns out to be significantly more efficient than that experienced with analogue functions from other \proglang{R} packages.
%% handling the PL mixture distribution. 

As possible directions of future work, the functions of the
\pkg{PLMIX} package could be further extended for the Bayesian
analysis of 
EPL mixtures, introduced by \cite{Mollica:Tardella} in the frequentist
domain, or to accommodate for the additional information provided by
subject- and/or item-specific covariates. Also 
we note that 
there is little availability of 
\proglang{R} routines to handle 
the presence 
% problem 
of ties in the ranking outcome and
% is neglected from a computational point of view with the 
this can stimulate a further improvement. Finally, visualization
techniques for the graphical 
% inspection 
illustration
of ranking data analysis could be also
included in a 
%future 
forthcoming version of the \pkg{PLMIX} package.

%\bibliographystyle{jss2}
%\bibliography{bcJSS}

 \newcommand{\noop}[1]{}

\end{document}